\documentclass[twocolumn]{aastex63}
\usepackage{graphicx,multirow, tabularx,chngcntr, appendix}

\newcommand{\romannumeralcap}[1]
    {\MakeUppercase{\romannumeral #1}}

\newcommand{\molh}{H$_2$}
\newcommand{\neii}{[Ne~{\sc ii}]}
\newcommand{\neiii}{[Ne~{\sc iii}]}
\newcommand{\msun}{$M_{\odot}$}
\newcommand{\ts}{TS16}
\newcommand{\mht}[1]{$M_{H2}$($T >$ #1)}
\newcommand{\tlprime}{$T_{l}^{'}$}

\newcolumntype{P}[1]{>{\centering\arraybackslash}p{#1}}

\received{XXX}
\revised{YYY}
\accepted{ZZZ}

\submitjournal{ApJ}

\shorttitle{M83 ISM Pt 1}
\shortauthors{Jones et al.}
\graphicspath{{./}{figures/}}

\begin{document}

\title{A {\it JWST}/MIRI View of the ISM in M83: \romannumeralcap{1}. Resolved Molecular Hydrogen Properties, Star Formation, and Feedback}

\correspondingauthor{Logan Jones}
\email{lojones@stsci.edu}

\author[0000-0002-1706-7370]{Logan H. Jones}
\affiliation{Space Telescope Science Institute, 3700 San Martin Drive, Baltimore, MD 21218, USA}

\author[0000-0003-4857-8699]{Svea Hernandez}
\affiliation{AURA for ESA, Space Telescope Science Institute, 3700 San Martin Drive, Baltimore, MD 21218, USA}

\author[0000-0002-0806-168X]{Linda J. Smith}
\affiliation{Space Telescope Science Institute, 3700 San Martin Drive, Baltimore, MD 21218, USA}

\author[0000-0001-5042-3421]{Aditya Togi}
\affiliation{Department of Physics, Texas State University, 601 University Drive, San Marcos, TX 78666, USA}

\author[0000-0003-0699-6083]{Tanio Diaz-Santos}
\affiliation{Institute of Astrophysics, Foundation for Research and Technology-Hellas (FORTH), Heraklion, 70013, Greece}
\affiliation{School of Sciences, European University Cyprus, Diogenes street, Engomi, 1516 Nicosia, Cyprus}

\author[0000-0003-4137-882X]{Alessandra Aloisi}
\affiliation{Space Telescope Science Institute, 3700 San Martin Drive, Baltimore, MD 21218, USA}

\author[0000-0003-2379-6518]{William Blair}
\affiliation{The William H. Miller III Department of Physics and Astronomy, Johns Hopkins University, 3400 N. Charles Street, Baltimore, MD 21218, USA}

\author[0000-0002-2954-8622]{Alec S. Hirschauer}
\affiliation{Space Telescope Science Institute, 3700 San Martin Drive, Baltimore, MD 21218, USA}

\author[0000-0001-9162-2371]{Leslie K. Hunt}
\affiliation{INAF--Osservatorio Astrofisico di Arcetri, Largo E. Fermi 5, I-50125 Firenze, Italy}

\author[0000-0003-4372-2006]{Bethan L. James}
\affiliation{AURA for ESA, Space Telescope Science Institute, 3700 San Martin Drive, Baltimore, MD 21218, USA}

\author[0000-0002-5320-2568]{Nimisha Kumari}
\affiliation{AURA for ESA, Space Telescope Science Institute, 3700 San Martin Drive, Baltimore, MD 21218, USA}

\author[0000-0002-7716-6223]{Vianney Lebouteiller}
\affiliation{AIM, CEA, CNRS, Université Paris-Saclay, Université Paris Diderot, Sorbonne Paris Cité, F-91191 Gif-sur-Yvette, France}

\author[0000-0003-2589-762X]{Matilde Mingozzi}
\affiliation{Space Telescope Science Institute, 3700 San Martin Drive, Baltimore, MD 21218, USA}

\author[0000-0002-9190-9986]{Lise Ramambason}
\affiliation{Institut fur Theoretische Astrophysik, Zentrum für Astronomie, Universität Heidelberg, Albert-Ueberle-Str. 2, D-69120 Heidelberg, Germany}

\begin{abstract}
    We present a spatially-resolved ($\sim$3 pc pix$^{-1}$) analysis of the distribution, kinematics, and excitation of warm \molh{} gas in the nuclear starburst region of M83. Our {\em JWST}/MIRI IFU spectroscopy reveals a clumpy reservoir of warm \molh{} ($>$200 K) with a mass of $\sim 2.3 \times 10^5$ \msun{} in the area covered by all four MRS channels. We additionally use the \neii{}~12.8~\micron{} and \neiii{}~15.5~\micron{} lines as tracers of the star formation rate, ionizing radiation hardness, and kinematics of the ionized ISM, finding tantalizing connections to the \molh{} properties and to the ages of the underlying stellar populations. Finally, qualitative comparisons to the trove of public, high-spatial-resolution multiwavelength data available on M83 shows that our MRS spectroscopy potentially traces all stages of the process of creating massive star clusters, from the embedded proto-cluster phase through the dispersion of ISM from stellar feedback.
\end{abstract}

\section{Introduction}
\label{sec:intro}

Since ongoing or imminent star formation primarily takes place in molecular clouds, scaling relations have been in use for decades \citep[e.g.,][]{Kennicutt98, Bigiel08, Bethermin23} to tie the star formation rate (SFR) to the mass of cold \molh{} gas in galaxies on both integrated and resolved scales. Once massive stars form and evolve, that same \molh{} gas is heated through a combination of far-ultraviolet (FUV) pumping and shocks from winds or supernovae, leading to strong emission from transitions in the mid-infrared \citep[MIR; e.g.,][]{Appleton06, Donnan23}. In other words, precise measurements of the mass and excitation state of \molh{} in galaxies -- especially at small scales relevant to cloud evolution and star formation --  are directly linked to both past and future generations of galaxy growth. However, because \molh{} molecules lack a permanent dipole moment, only much weaker quadrupole (ro-vibrational) transitions are permitted, with allowed upper energy states starting at $E / k$ = 510 K.

Central bursts of star formation (SF) in massive galaxies are excellent laboratories for understanding the interwoven connections between \molh{} gas, SF, and feedback episodes and their impact on the evolution of galaxies. The {\em Spitzer} era of space-based infrared astronomy revealed a rich suite of spectroscopic features to diagnose the physical state of the interstellar medium (ISM) in SF galaxies -- for example, emission from polycyclic aromatic hydrocarbon (PAH) grains \citep{Maragkoudakis18}, or line emission from shocked and/or photoionized atomic and molecular gas \citep[][]{Inami13}.  With an order of magnitude improvement in sensitivity and in spatial and spectral resolution, the advent of the Medium Resolution Spectrometer (MRS) on {\em JWST}/MIRI has further broadened the diagnostic space for understanding SF and feedback at previously unattainable scales \citep{Young23, Bik24, Rigo24}.

As the nearest face-on spiral galaxy with a nuclear starburst ($D$ = 4.6 Mpc, \citealt{Saha06}), M83 has been studied extensively across the electromagnetic spectrum, including the X-ray \citep[e.g.,][]{Hunt21, Wang21}, UV \citep[e.g.,][]{Thilker05, James14, Hernandez21}, optical and near-IR \citep[e.g.,][]{Mast06, Knapen10, Winkler23}, and millimeter to radio \citep[e.g.,][]{Russell20, Koda23} regimes. Additionally, the nucleus and part of the disk of M83 was recently observed with wide- and medium-band filters from {\em JWST}/NIRCam and {\em JWST}/MIRI as part of the FEAST program (GO 1783, PI A.\ Adamo; Adamo et al., in prep.), which provide unprecedented views of the bulk stellar populations, sites of embedded SF, and diffuse emission from PAHs. Finally, M83 contains numerous young, massive star clusters (YMCs) with archival FUV spectroscopy from {\em HST}/COS. The COS spectra provide direct constraints on the intensity and hardness of the ambient ionizing radiation field and of the amount of mechanical feedback from stellar winds (via, e.g. C{\sc IV} $\lambda\lambda$ 1549,1551 \AA{}), each of which acts to regulate the excitation of nearby molecular clouds.

In \citet{Hernandez23} (hereafter H23), we reported top-level results from {\em JWST} GO program 02219 (PI S.\ Hernandez), in which we investigated the spatially-integrated spectral properties of our four MRS pointings covering the central $\sim$200$\times$200 pc$^2$ of M83's nuclear region, including the optical nucleus. These spectra provided insights into the ISM conditions of M83's center at the scale of $\sim$100 pc, as well as how they vary from region to region. In this work, we return to the same dataset to inspect the properties of the warm \molh{} gas at the level of individual spaxels (0\farcs13 or $\sim$3 pc). The high spatial resolution and sensitivity of these data allows us to search for quantitative and qualitative changes in the gas properties of the molecular and ionized gas with respect to the location of the optical nucleus and (un)obscured YMCs.

We structure this paper as follows. In Section \ref{sec:obs}, we describe the observations and data reduction steps. In Section \ref{sec:methods}, we outline the spectral fitting tools and procedures we use to fit our IFU data on a spaxel-by-spaxel basis, followed by a summary in Section \ref{sec:analysis} of our methods for determining \molh{} excitation, SFR, and other physical properties. We put our findings for the warm \molh{} content into a broader multiwavelength context in Section \ref{sec:disc}, including a qualitative discussion of the different stages of stellar feedback and ISM clearing that M83 displays in its nuclear region.

\section{Observations}
\label{sec:obs}

We use the MIRI/MRS observations from program GO 02219 (PI: Hernandez), which were obtained between 2022 July 5 -- 9. The raw MIRI/MRS files were retrieved from the Mikulski Archive for Space Telescopes (MAST); all JWST data used in this paper can be found in MAST:10.17909/a61h-f081. Additional details on the observation strategy are given in H23, though we summarize the most important points below.

\begin{figure*}[t]
\centering
\includegraphics[width=0.9\textwidth,trim={0.25cm, 0.1cm, 0.2cm, 0.1cm}, clip]{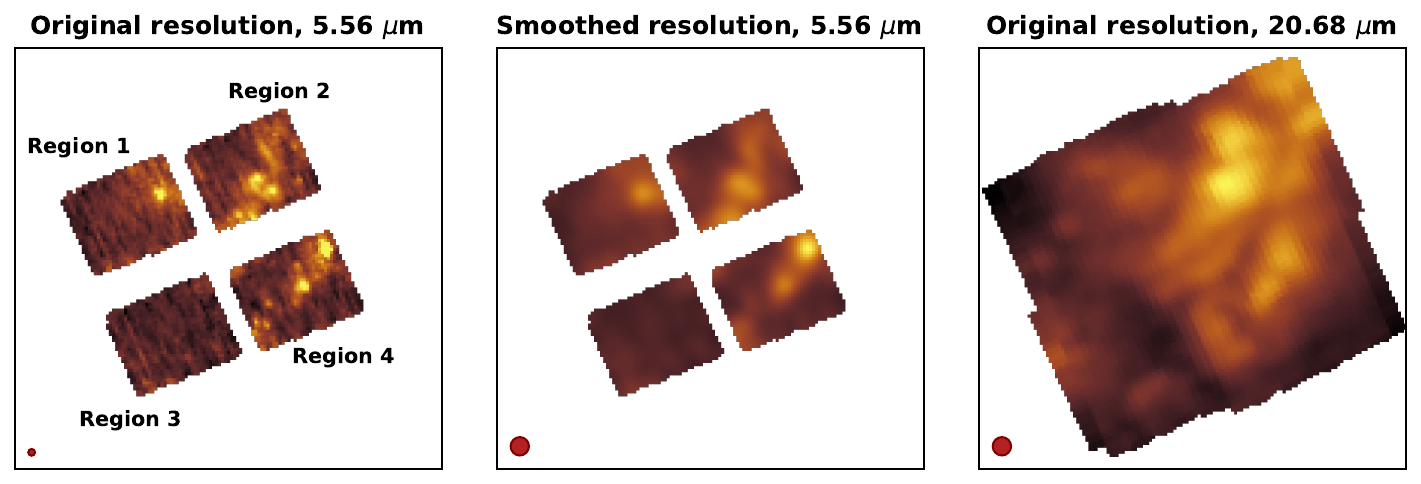}
\caption{
An illustration of the Gaussian convolution process we use to smooth our IFU observations of M83's nuclear region to a common spatial resolution. {\em Left:} A continuum-only wavelength slice at 5.56 \micron{} at its native resolution (PSF FWHM 0.29\arcsec), with the four Regions labeled. {\em Center:} The same Ch1 wavelength slice, now convolved with a Gaussian to a PSF FWHM of 0.788\arcsec.  {\em Right:} The reddest wavelength slice in our truncated IFU cube (20.68 \micron) at its native resolution (PSF FWHM 0.788\arcsec). The rightmost image is displayed in log scale to better show faint structures.
\looseness = -2
}
\label{fig:convolution}
\end{figure*}

The primary MRS observations consist of four pointings in a 2 $\times$ 2 mosaic, with a 4-point dither pattern applied to each pointing to achieve optimal sampling throughout the MRS field-of-view (FOV) and to mitigate detector artifacts. This results in contiguous spatial coverage across the mosaic in MRS channels 3 and 4 but not channels 1 and 2. The specific pointings (labelled as Regions 1, 2, 3, and 4 in the left panel of Figure \ref{fig:convolution} and in select figures thereafter) were strategically chosen to provide full spatial coverage of the YMCs in all four MRS channels, with the FOV per pointing increasing from 3.2\arcsec $\times$ 3.7\arcsec{} in Channel 1 (4.9 --  7.65 \micron) to 6.6\arcsec $\times$ 7.7\arcsec{} in Channel 4 (17.7 -- 27.9 \micron). The observations were obtained over three visits, with one visit each dedicated to observing with the SHORT (A), MEDIUM (B), and LONG (C) gratings. The total integration time per pointing+grating was 2264 s. 

Due to the extended nature of M83's nuclear region, we required a dedicated sky observation to accurately measure and correct for the thermal background. We thus also obtained a background pointing with MRS along a sightline well off the disk of M83, with integration times equal to those of the primary pointings. All MRS observations were reduced with the standard {\em JWST} pipeline (version 1.10.2), with the spatial pixels gridded to a common scale of 0.13\arcsec{} pix$^{-1}$ and the output type set to {\tt multi} in the Spec3 pipeline to automate the spectral stitching between bands/channels. We also applied residual fringe corrections available in the {\em JWST} pipeline to both the three- dimensional and one-dimensional observations, though as we discuss below, significant fringing is present at the longest wavelengths even with these corrections. 

In the course of reducing and analyzing these data, we also noted significant discontinuities in the continua of some individual spaxel spectra. These discontinuities are most prominent in the Channel 4 data and occur approximately at the wavelengths where the spectral coverage begins or ends for the 4A, 4B, and 4C bands, which we attribute to issues with flux calibration across bands within the MRS pipeline. These artificial breaks in the continuum flux are often severe enough to affect the quality of our fits to the warm dust continuum at $\lambda_0 \gtrsim 20$ \micron{} (see Section \ref{sec:methods}), which in turn complicated our emission line extraction and fitting. We opted to truncate the wavelength axis of our combined data cube at 20.68 \micron{} (the blue limit of band 4B) to mitigate the impact of the discontinuities on our full spectral fitting. For this work, we also restrict ourselves only to the parts of our IFU data cube covered by MRS channel 1 (the smallest FOV of the four channels) to ensure contiguous spectral coverage down to $\sim$5 \micron.

Finally, we spatially smoothed our data cube to a common resolution of 0.788\arcsec{} FWHM ($\sim$18 pc at $D = 4.6 Mpc$), which is the PSF FWHM of MRS at our long-wavelength cutoff of 20.68 \micron{} (see Equation 1 of \citealt{Law23}). For each wavelength slice in the IFU cube, we determine the quadrature difference between the PSF FWHM at 20.68 \micron{} and the nominal PSF FWHM at that wavelength (again according to Equation 1 of \citealt{Law23}), and then convolve the slice with a 2D Gaussian with the corresponding width, $\sigma = \sqrt{\sigma_{20.68}^{2} - \sigma_{\lambda}^{2}}$. We show an example of this process in Figure \ref{fig:convolution}. 

\section{Methods}
\label{sec:methods}

\subsection{PAHs, Dust Continuum, and Attenuation}
\label{subsec:methods:att}

\begin{figure*}[t]
\centering
\includegraphics[width=0.45\linewidth,trim={0.22cm, 0.1cm, 0.77cm, 0.1cm}, clip]{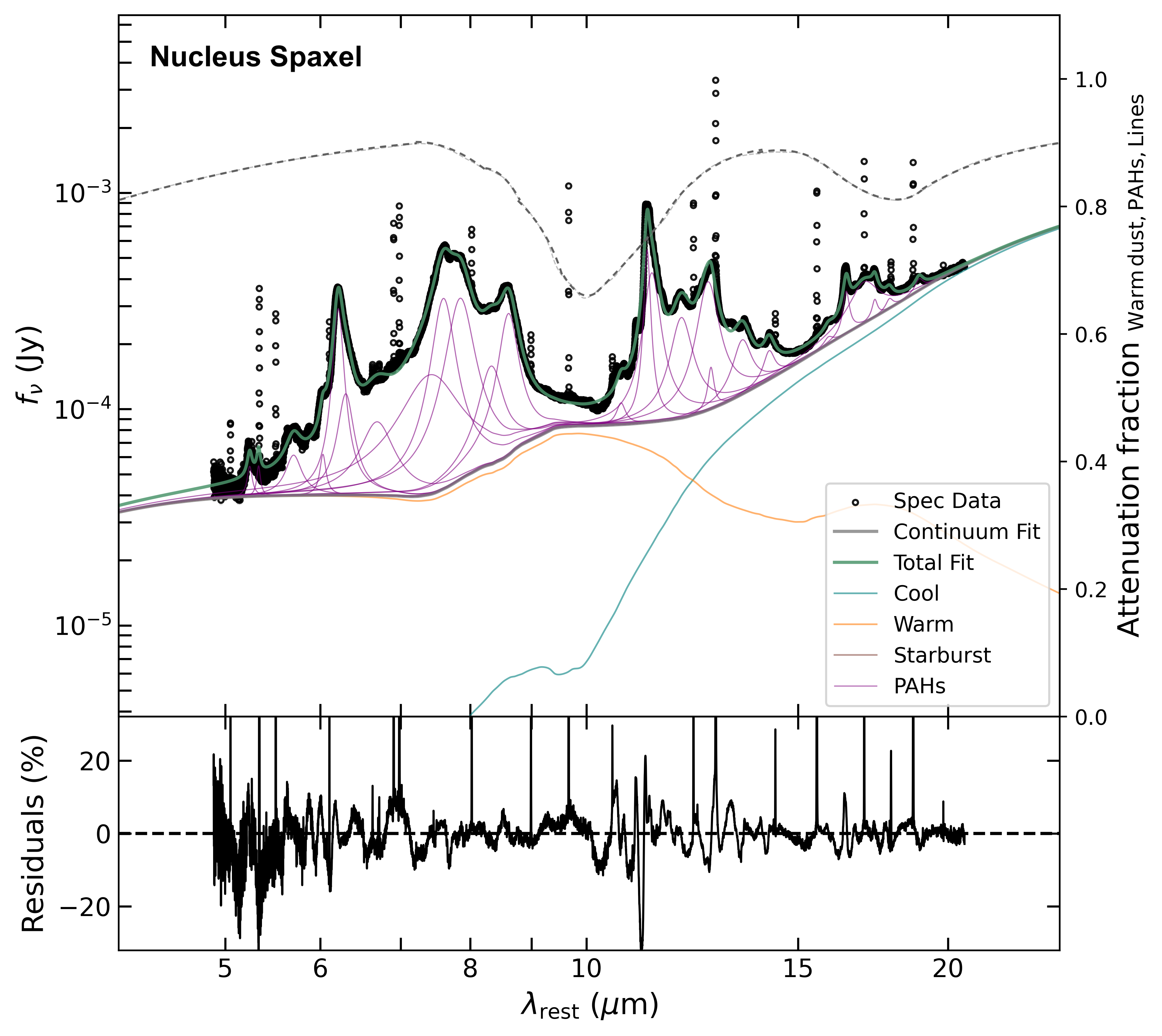} \hspace{8pt}
\includegraphics[width=0.45\linewidth,trim={0.79cm, 0.1cm, 0.2cm, 0.1cm}, clip]{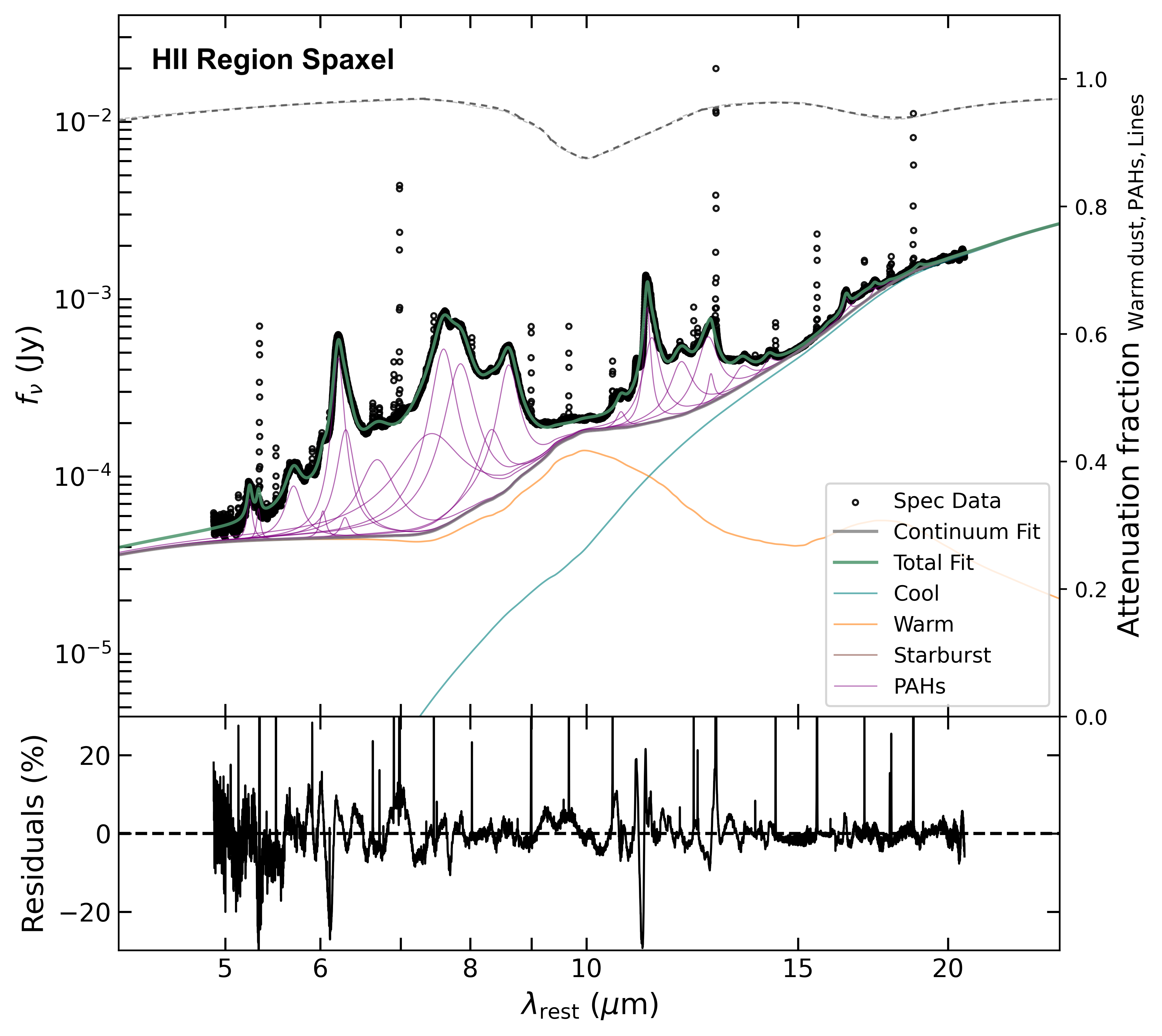} 
\caption{
Example best fits (top) and residuals (bottom) from {\tt CAFE} for two spaxel spectra in our MRS cube, which illustrate the diversity of dust conditions across our FOV. Model PAH features and continuum emission from cool and/or warm dust ($\sim$60 and 200 K) are plotted as marked in the legend. The dotted-grey curve shows the attenuation fraction as a function of wavelength. 
\looseness = -2
}
\label{fig:cafefits}
\end{figure*}

\begin{figure}[t]
\centering
\includegraphics[width=\linewidth,trim={0.2cm, 0.1cm, 0.2cm, 0.1cm}, clip]{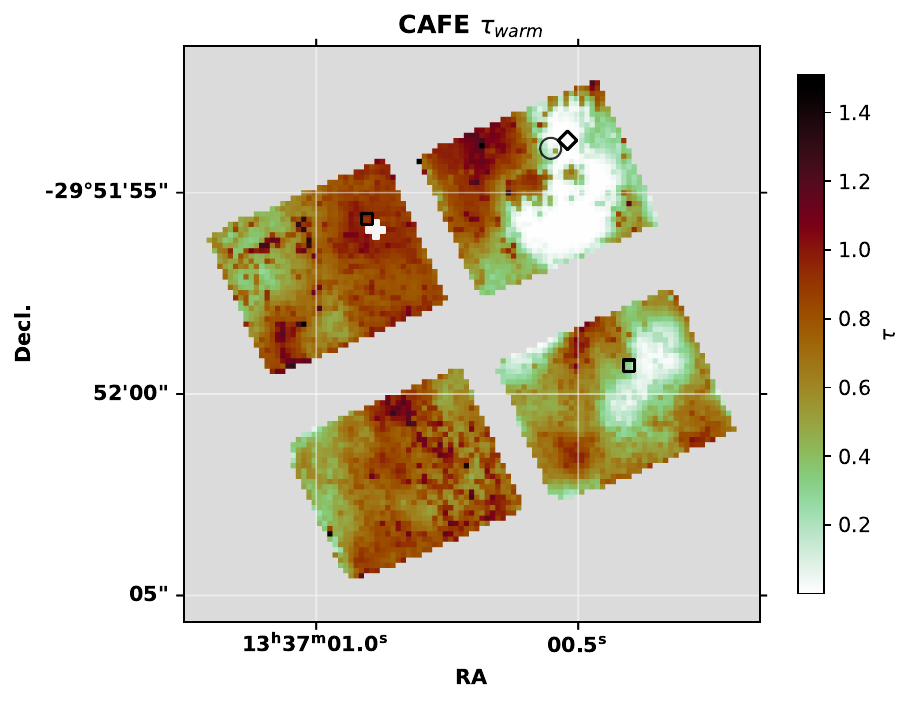}
\caption{
Spatial distribution of best-fit values of $\tau_{warm}$ from {\tt CAFE}, which indicates the amount of attenuation applied to PAHs and emission lines in the {\tt CAFE} framework. The locations of the single-pixel spectra and {\tt CAFE} fits shown in Figure \ref{fig:cafefits} are marked with black squares.
\looseness = -2
}
\label{fig:cafetau}
\end{figure}

We use the Continuum and Feature Extraction software ({\tt CAFE}; \citealt{cafe}; Diaz-Santos et al., in prep.)\footnote{https://github.com/GOALS-survey/CAFE} to model PAH features, dust continuum emission, and extinction in our spaxel spectra. We began by assuming a systemic redshift of $z$ = 0.001733 \citep{MeyerHipass} to shift our MRS cube into the approximate rest frame for fitting with {\tt CAFE}. We used a modified version of the MIRI-optimized starburst parameter file in {\tt CAFE}, which models the mid-IR continuum as the sum of reprocessed dust continua and ``starburst" continua. The stellar model consists of a mixture of 2, 10, and 100 Myr-old single stellar population (SSP) templates from Starburst99 \citep{starburst99}. However, we note that only $\sim$half of spaxels (mostly concentrated near the optical nucleus and generally dust-free areas in Regions 3 and 4) ultimately include a non-zero stellar component in their best-fit {\tt CAFE} model, with the 10 Myr component dominant.

Our primary dust model includes emission and absorption from cool (40 -- 100 K) and warm (100 -- 400 K) dust and assumes a screen geometry. However, we found that the two-component dust model produced physically implausible fits (i.e., optical depths $\gtrsim$10 at 14 \micron) for some highly-obscured areas in Region 2 whose continuum spectra were dominated by emission from cool dust. For these $\sim$70 spaxels, we instead fitted the spectra with a single dust component with temperatures between 40 -- 100 K.
Finally, we assume a Galactic interstellar radiation field as the illuminating source for all model dust components. 

Individual PAH features (drawn from a feature library included in {\tt CAFE}) were fitted with Drude profiles. We allow the centroid of the Drude profiles to vary by up to 500 km s$^{-1}$, which is sufficient to account for (1) uncertainties in the MIRI/MRS wavelength calibration \citep{Rigby23} and (2) velocity shifts due to the kinematics of the PAH-bearing ISM in M83. We also allow the width of each component to vary by up to $\pm$20\% of their default value in {\tt CAFE}.  We did not fit emission lines simultaneously with PAHs and mid-IR continua for the sake of computational efficiency, as this would have nearly tripled the number of fittable model parameters for a single spaxel, and instead perform our own line fitting after subtracting the best-fit dust+PAH model as described in the next section. Finally, we assumed the \citet{Ossenkopf92} extinction curve, which includes strong silicate absorption features at 9.7 and 18 \micron.

We show representative spaxel spectra and {\tt CAFE} fits in Figure \ref{fig:cafefits}, which illustrate the diversity of dust properties in our data. The left panel shows a spectrum toward the mildly-obscured optical nucleus (($\alpha$, $\delta$) = (13:17:0.887, -29:51:55.921) as measured from our MRS data), while the right panel shows a spectrum toward a UV-bright star cluster with partially-cleared foreground ISM. {\tt CAFE} produces acceptable model fits across the wavelength range we consider in this work (with residuals of $\pm$5\% in line-free regions), with the exception of the highly structured PAH complexes at 6 -- 6.6 and 11 -- 13 \micron, where the fitting algorithm sometimes fails to identify weak features hidden in the broad wings of more dominant PAH features. Finally, we note that the fractional attenuation (dotted grey curve) experienced by PAH features and emission lines can be significant, even in the mid-IR. For the nuclear spaxel spectrum, for example, roughly 10\% of the light at 15 \micron{} is attenuated; this increases to 30\% at 10 \micron, driven by the deep silicate absorption feature at 9.7 \micron.

In Figure \ref{fig:cafetau}, we show the spatial distribution of the optical depth due to warm dust at 14 \micron, $\tau_{warm}$, as inferred by {\tt CAFE} across the FOV of our MRS cube. Although the median optical depth across our FOV is fairly low ($\tau_{warm} \sim$ 0.64), we note regions where the dust is optically thick ($\tau_{warm} \gtrsim$ 1), particularly in Regions 1 and 2. Because the amount of attenuation experienced by PAHs and emission lines in {\tt CAFE} is tied to the optical depth of warm dust, this means that all line fluxes in these regions will be boosted by up to $\sim$40\%.

\subsection{Emission Line Maps}
\label{subsec:methods:emlines}

After obtaining reliable PAH+continuum models and attenuation estimates across the FOV of our combined cube using the fitting procedure outlined above, we turn to the fitting of emission lines. We perform a local continuum fitting and subtraction in the vicinity of the lines analyzed in this work.
We then use {\tt specutils} to fit a single Gaussian to each of the \molh{} transitions listed in Table \ref{tab:emlist}. We also perform single and double Gaussian fits to the \neii{} 12.81 \micron{} and \neiii{} 15.56 \micron{} lines (see Section \ref{subsec:analysis:sfr}, which we use as an SFR diagnostic and approximate tracer of the hardness of the ambient UV field). All fits were allowed to vary independently (i.e., the kinematics of the \molh{} lines are not tied). We integrate over the best-fit Gaussian profile to obtain total line or line component fluxes in units of erg s$^{-1}$ cm$^{-2}$. 

We note that a small but significant number of spaxels (particularly those near the edge of the Ch1 FOV) have no data in certain narrow wavelength intervals, most of which lie below 8 \micron; we attribute this to a combination of the dither pattern and the spectral stitching routines in the MRS pipeline, which results in edge spaxels having sporadic wavelength coverage when combining our data into a single cube. Unfortunately, these missing wavelength intervals frequently overlap with the expected positions of the S(5) -- S(8) transitions of \molh, which limits the number of transitions that go into our \molh{} excitation modeling near the edge of the Ch1 FOV.

We estimate the significance of a line detection by estimating the noise in a nearby, line-free region of the continuum-subtracted spectra, in a manner similar to that used in  \citet{DelPino23}. Specifically, we measure the RMS flux density in 0.03 \micron{}-wide windows in the pseudo-continuum around each line, then multiply this value by 2$\times$ the FWHM of the line as an estimate of the amount of ``signal" that would be present in the absence of any line emission. Unless otherwise specified, we adopt a threshold of S/N $>3$ for all lines and line components, in line with other emission-line studies \citep[e.g.,][]{Mingozzi22}.

Finally, we correct all line fluxes and uncertainty estimates using the wavelength-dependent attenuation fraction inferred by {\tt CAFE} for each spaxel, shown as a dotted-grey line in Figure \ref{fig:cafefits}. In other words, at the rest-frame wavelength of each line, we extract the value $A$ of the grey curve \citep{Ossenkopf92} and multiply the observed flux and flux uncertainty by (1/$A$) to recover the intrinsic, de-reddened line flux without artificially inflating its S/N.

\begin{table}[tb]
    \caption{{\sc~List of \molh{} Rotational Transitions}}
    \vspace{-0.3cm}
    \label{tab:emlist}
    \begin{center}
    \begin{tabular}[t]{cccc}
        \hline
        \hline
        Transition & Name & $\lambda_0$ & $E_u / k$\\
        ($\nu = 0$) &   & (\micron) & (K) \\
        \hline
       $ J = 9 \rightarrow$ 7 & S(7) & 5.5115 & 7197 \\
       $ J = 8 \rightarrow$ 6 & S(6) & 6.1088 & 4830\\
       $ J = 7 \rightarrow$ 5 & S(5) & 6.9095 & 4586 \\
       $ J = 6 \rightarrow$ 4 & S(4) & 8.0250 & 3475 \\
       $ J = 5 \rightarrow$ 3 & S(3) & 9.6649 & 2504 \\
       $ J = 4 \rightarrow$ 2 & S(2) & 12.2786 & 1682 \\
       $ J = 3 \rightarrow$ 1 & S(1) & 17.0348 & 1015 \\
        \hline
    \end{tabular}
    \end{center}
\end{table}

\section{Analysis}
\label{sec:analysis}

We show maps of the attenuation-corrected flux, gas velocity, and line FWHM for the S(1), S(3), S(5), and S(7) rotational \molh{} lines in Figure \ref{fig:h2vis}. Velocity maps are shown with respect to the systemic redshift, $z$ = 0.001733, corresponding to a recession velocity of 519 km s$^{-1}$ \citep{MeyerHipass}. Line FWHMs have been corrected for instrumental broadening, using Equation 1 of \citet{OJones23} to determine the spectral resolution at the wavelength of each line except for S(1). Their Equation 1 underestimates the observed resolving power of MRS at $\sim$15 -- 17 \micron{} (see their Figure 5 and the {\em JWST} user documentation\footnote{https://jwst-docs.stsci.edu/jwst-mid-infrared-instrument/miri-observing-modes/miri-medium-resolution-spectroscopy}), leading us to adopt $R$ = 3300 for the S(1) transition and other lines in this wavelength interval.

The flux and kinematic maps of these transitions show generally similar structures, with prominent \molh{} line emission from the optical nucleus (cross symbol in Figure \ref{fig:h2vis} and thereafter), highly-extincted knots in Region 2, and other clumps centered around a string of UV-bright YMCs in Region 4. However, we note that higher-excitation lines are stronger (relative to S(1)) in the optical nucleus and weak or undetected in much of Regions 3 and 4. The lines are generally resolved (corrected FWHMs $\gtrsim$30 km s$^{-1}$) across much of our FOV, with some filamentary or clumpy structures reaching up to FWHM $\sim$140 km s$^{-1}$. We note that the portions of Region 3 which are undetected in higher-excitation lines are coincident with one such patch of highly dispersed gas, as observed in the lower-excitation lines. We also identify a kinematically distinct point source in Region 2, the position of which is marked with a circle in the first and final rows of Figure \ref{fig:h2vis}. We further discuss the possible nature of this object in Section \ref{subsec:disc:baby}, though we note that it is {\em only} detected robustly in the flux and kinematics of S(1) and weakly in S(2). No such structure is seen at this position in \neii, \neiii, or other \molh{} lines.

\subsection{\molh{} Excitation Modeling}
\label{subsec:analysis:h2ex}

\begin{figure*}[t]
\centering
\includegraphics[width=0.95\linewidth,trim={0.25cm, 0.1cm, 0.3cm, 0.1cm}, clip]{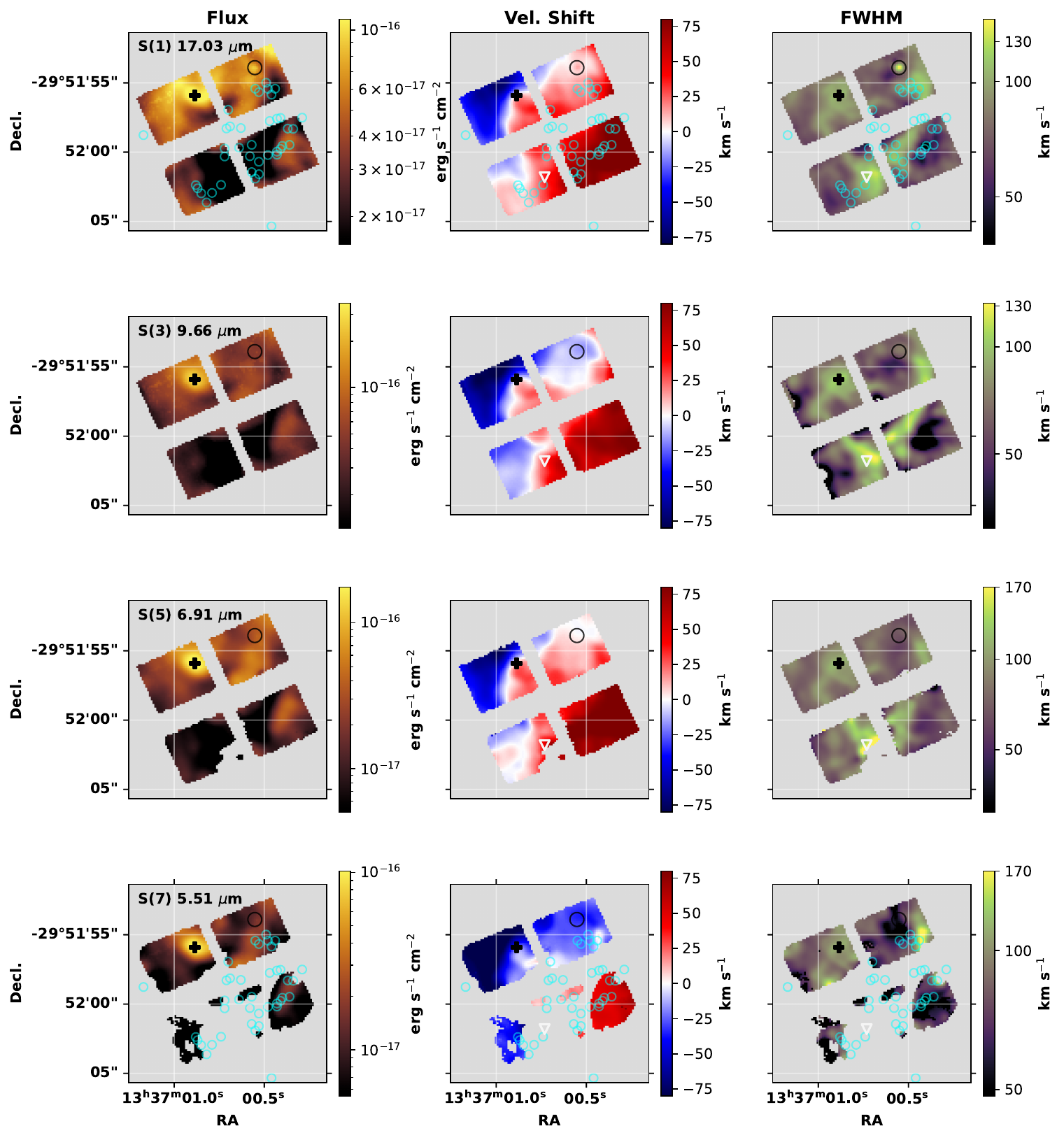}
\caption{
Maps of attenuation-corrected line flux ({\em left}), velocity offsets from systemic ({\em center}), and line FWHM ({\em right}) for odd-$J$ pure rotational transitions of \molh{} in our MRS cube. All lines are required to have S/N $> 3$ to be included in the maps and our \molh{} excitation fitting. Cyan circles (0\farcs6 diameter) mark the position of UV-bright star clusters from \citet{Harris01}. The plus symbol, circle, and downward triangle respectively note the position of the optical nucleus, an unidentified \molh{} point source (see Section \ref{subsec:disc:baby}), and an unidentified point source showing [Ne~{\sc v}] 14.32 \micron{} and [Ne~{\sc vi}] 7.65 \micron{} emission (see Section \ref{subsec:disc:interestingplaces}).
\looseness = -2
}
\label{fig:h2vis}
\end{figure*}

To obtain an estimate of the warm \molh{} mass per spaxel, we use the \molh{} excitation model of \citet{Togi16} (hereafter \ts). In brief, the model assumes that the rotational temperatures of \molh{} gas follow a power-law distribution $dN = mT^{-n}dT$ for column density $N$, temperture $T$, and a constant $m$. Given the analytical expressions in \ts{} for $m$ and other relevant quantities (e.g., column density of \molh{} molecules in the upper energy level $j$ $N_j$), the model parameters that must be constrained by the data are the slope $n$ and the upper and lower bounds of the temperature distribution $T_u$ and $T_l$. \ts{} find that the model fit quality and inferred \molh{} mass become insensitive to the choice of $T_u$ at temperatures above 2000 K and thus assume a fixed $T_u$ = 2000 K, which we also adopt in this work. $n$ and $T_l$ can then be constrained directly from the data by minimizing the difference between the {\em observed} values of $N_j$ (which are tied to the line fluxes) and the {\em predicted} values of $N_j$ for some $n$ and $T_l$. Both the observed and predicted columns are normalized to the S(1) line for the fitting procedure.

For this work we created a Python adaptation of the \ts{} model, which we release for community use\footnote{Link TBD}. Our Python version yields best-fit values of $n$, $T_l$, and \molh{} masses $M (T > T_l)$ that are identical to the original implementation in IDL to within $< 1\%$, which we confirm through tests using the \ts{} sample. Parameter uncertainties on $n$ and $T_l$ are slightly higher in our Python implementation, which we attribute to the choice of minimization algorithm (the Levenberg-Marquardt algorithm in IDL vs.\ the Trust Region Reflective algorithm in {\tt scipy}'s {\tt curve\_fit} function).

\begin{figure}[t]
\centering
\includegraphics[width=\linewidth,trim={0.2cm, 0.1cm, 0.2cm, 0.1cm}, clip]{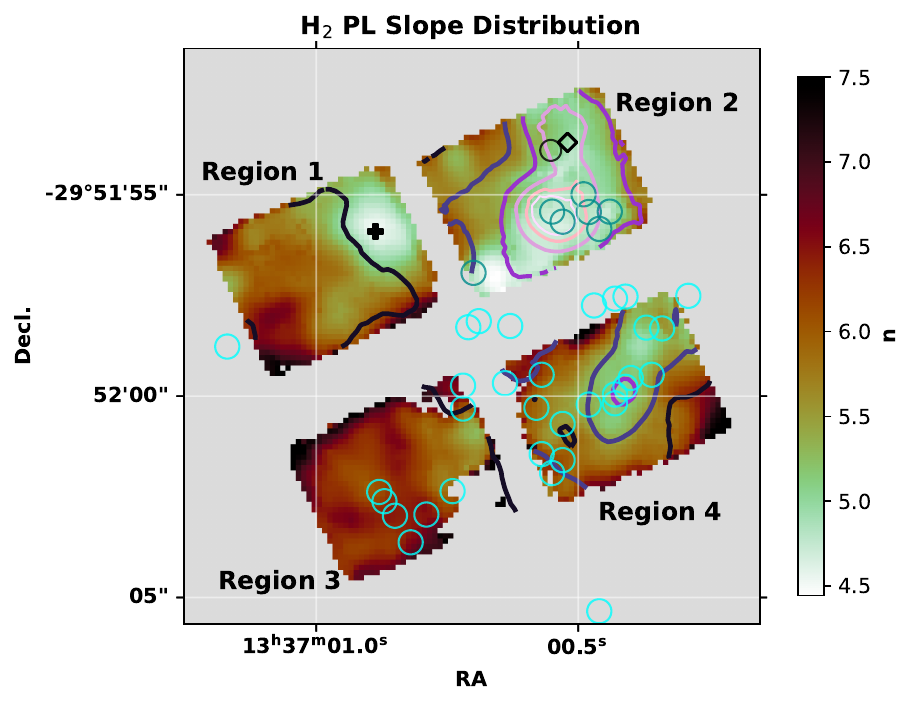}
\includegraphics[width=\linewidth,trim={0.2cm, 0.1cm, 0.2cm, 0.1cm}, clip]{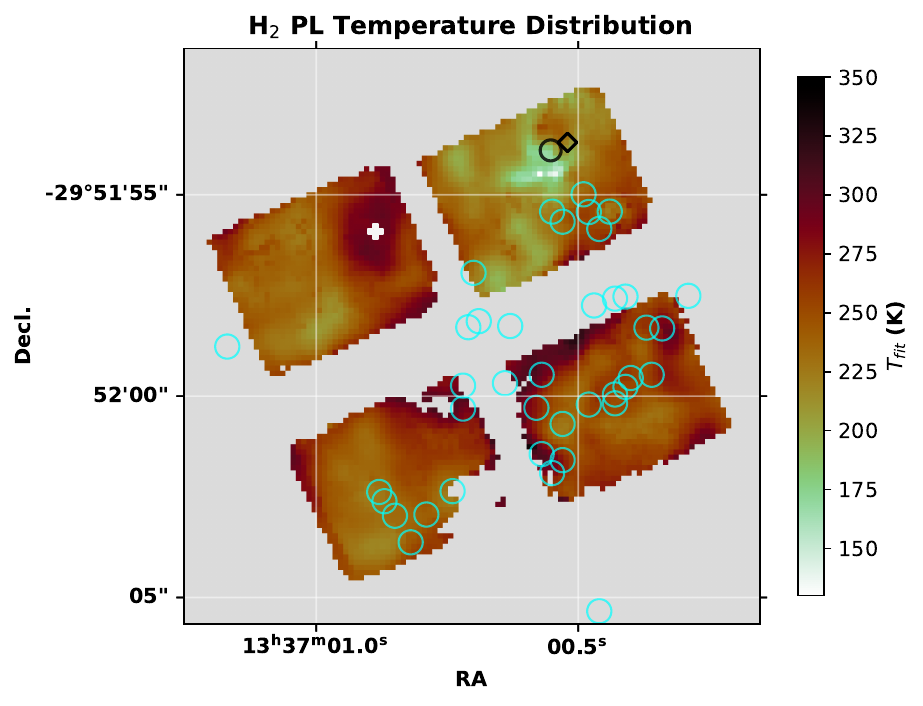}
\caption{
Spatial distribution of best-fit power-law slopes ({\em top}) and lower cutoff temperature $T_{l, fit}$ ({\em bottom}) for the continuous-temperature power-law model of \ts{}. Blue circles (0\farcs6 diameter) mark the position of UV-bright star clusters from \citet{Harris01}. The plus symbol, diamond, and circle respectively note the position of the optical nucleus, a partially-embedded YMC, and an unidentified \molh{} point source described in Section \ref{sec:analysis}. Purple-pink contours trace $\Sigma_{SFR}$ using the \citet{Zhuang19} diagnostic at intervals of [2, 5, 15, 30, 45, 55] $M_{\odot}$ yr$^{-1}$ kpc$^{-2}$ (see Section \ref{subsec:analysis:sfr})
\looseness = -2
}
\label{fig:h2slopes}
\end{figure}

We show the map of best-fit slopes $n$ in Figure \ref{fig:h2slopes}. 
For the remainder of our resolved \molh{} analysis, we consider only those pixels with four or more \molh{} line detections and with relative error $\delta n / n < 25\%$ to mitigate the impact of poor power-law fits on our total \molh{} mass estimates. This mostly affects pixels along the FOV edges and the corners of Region 3, which is generally very faint or undetected in \molh{} transitions beyond S(3), as seen in Figure \ref{fig:h2vis}. The median slope and 1$\sigma$ range across our FOV, $n = 5.75 \pm 0.64$,
reaches the upper end of the 3.79 $< n < $ 6.39 range reported in \ts{} for their sample of SINGS galaxies and is somewhat steeper than (but consistent with) their sample mean of $n = 4.84 \pm 0.61$. This indicates, on average, relatively little \molh{} in highly excited states across our four regions in M83, though with significant variation. We note quite shallow slopes (i.e., relatively high \molh{} excitation) are clustered near the optical nucleus and the clumps or strings of UV-bright star clusters in Regions 2 and 4, while the steepest slopes are concentrated in Region 3. 
We also find that the best-fit values of $T_l$ across our FOV vary from $\sim$130 -- 343 K (median 247 K), with the lowest values of $T_l$ coincident with CO-bright knots in Region 2 (see the discussion in Section \ref{sec:disc}) and the highest found near the optical nucleus. 

\subsection{Warm \molh{} Mass Estimates}
\label{subsec:analysis:h2mass}

With the power-law \molh{} temperature distribution thus constrained, we can then use Equations 9 and 14 in \ts{} to derive total \molh{} column densities and masses per spaxel. However, while the slope of the temperature distribution in each spaxel is reasonably well-constrained by the data, there remain several justifiable choices for the lower temperature bound \tlprime{} for the column integral. Since the bulk of the cold, imminently star-forming \molh{} cannot be probed through rotational emission, it is natural to wonder if extrapolation of the power-law model to lower temperatures will produce physically sensible results. A proper choice for \tlprime{} is particularly critical for our data given the steep slopes we measure in some areas, as extrapolating to lower and lower values of \tlprime{} may cause the total inferred \molh{} mass to balloon beyond what is inferred through CO and other tracers.

\begin{figure}[t]
\centering
\includegraphics[width=\linewidth,trim={0.2cm, 0.1cm, 0.2cm, 0.1cm}, clip]{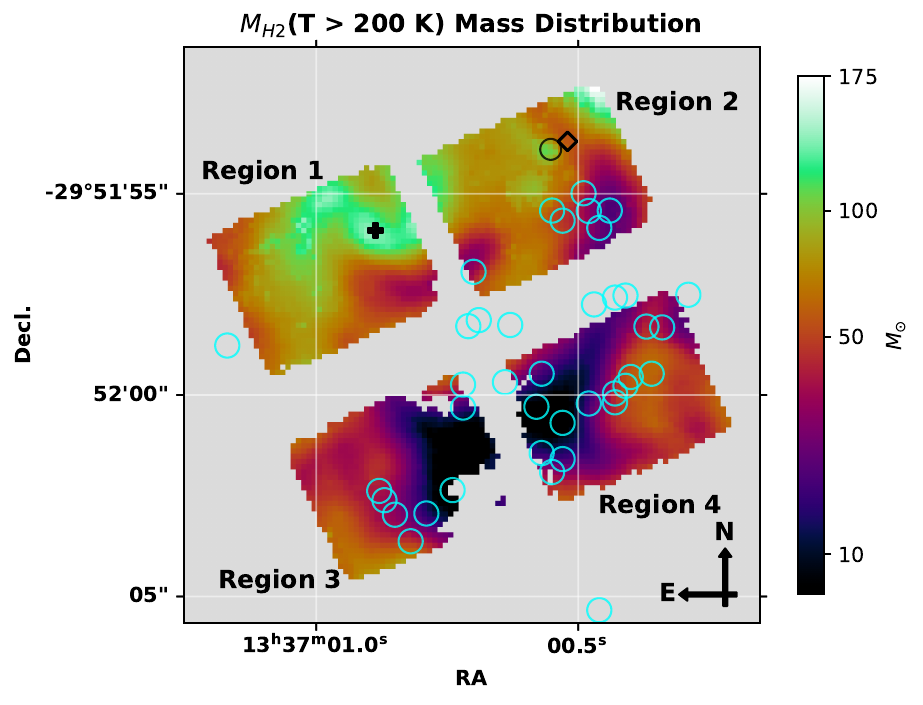}
\includegraphics[width=\linewidth,trim={0.2cm, 0.1cm, 0.2cm, 0.1cm}, clip]{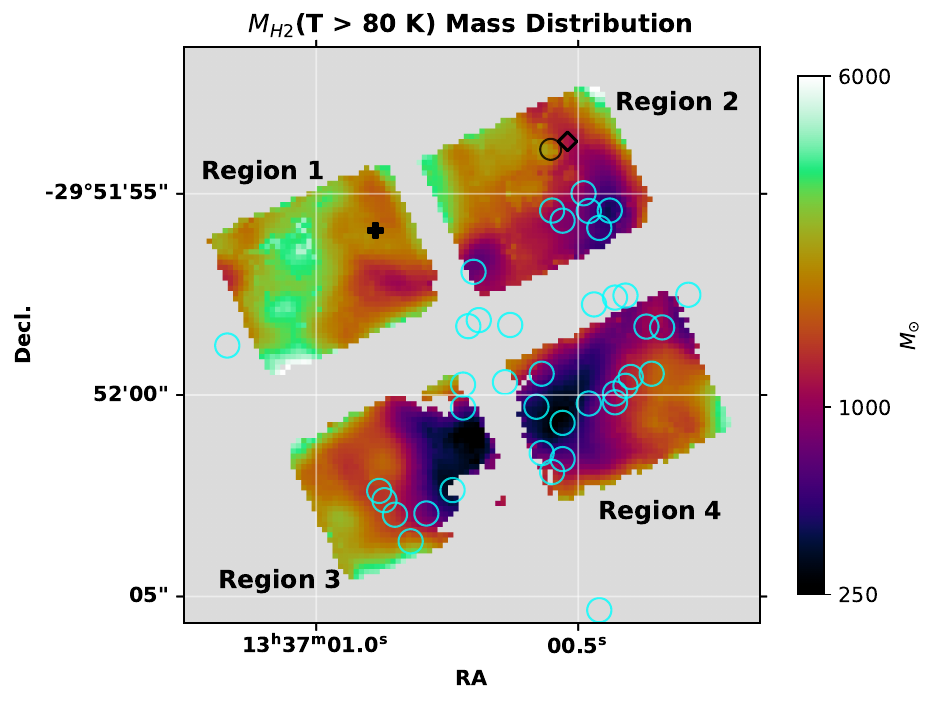}
\caption{
Mass per 0\farcs13 spaxel of \molh{} as estimated from the power-law model of \ts, integrated to a common \tlprime{} = 200 K ({\em top}) and 80 K ({\em bottom}). Cyan circles (0\farcs6 diameter) mark the position of UV-bright star clusters from \citet{Harris01}. Other symbols are as defined in Figure \ref{fig:h2slopes}. 
\looseness = -2
}
\label{fig:h2_warmmass}
\end{figure}

Using the best-fit power-law slopes per spaxel, we therefore construct three different maps of the \molh{} mass distribution, each with a different definition of \tlprime:

\begin{enumerate}
  \item \tlprime{} is set equal to the best-fit $T_l$ for each spaxel. This approach provides an estimate of the {\em warm} \molh{} that is the most directly constrained by the data.
  \item \tlprime{} is fixed at 200 K across the FOV, which imposes a modest extrapolation of the temperature distribution for most spaxels but provides a more uniform estimate of the warm \molh{} than (1). 
  \item \tlprime{} is fixed at 80 K across the FOV, which is a typical temperature below which the \molh{} reservoir can change in mass without measurable changes to the rotational line fluxes in the {\em Spitzer}/IRS spectroscopy analyzed in \ts. 
\end{enumerate}

Unless otherwise specified, we propagate only the uncertainty on $n$ (and not $T_l$) to the gas mass estimates. Adopting the first two definitions of \tlprime, the warm \molh{} mass across our FOV ranges between (1) $1.50^{+0.09}_{-0.09} \times 10^5 M_{\odot}$ for a variable \tlprime{} and (2) $2.26^{+0.17}_{-0.19} \times 10^5 M_{\odot}$ for \tlprime{} = 200 K. Adopting the third definition with \tlprime{} = 80 K, which taps into the cooler gas, yields (3) $7.93^{+3.52}_{-2.59} \times 10^6 M_{\odot}$. In Figure \ref{fig:h2_warmmass}, we show the mass per 0\farcs13 spaxel inferred using definitions (2) and (3) -- in other words, maps of the warm and cool+warm \molh{} gas over our FOV.

We note that this work is one of the first uses of the \ts{} model on the $\lesssim$5 pc scale, compared to the $\sim$kpc-scale regions probed by {\em Spitzer}/IRS in the sample analyzed in \ts. Even in the era of {\em JWST}, other recent works that follow \ts{} to model \molh{} masses and excitation from MRS spectroscopy only probe down to scales $\sim$100 pc \citep[e.g.,][]{Lai22,Armus23, Zhang23, Costa24}. To assess the importance of physical scale on the derived mass estimates, we applied {\tt CAFE} and our custom line-fitting procedure to the integrated-light spectrum (created through the coaddition of every spaxel in our common FOV, which approximates the total spectrum of the central 200$\times$200 pc$^2$ of M83).

\begin{figure*}[t]
\centering
\includegraphics[width=0.95\linewidth,trim={0.2cm, 0.1cm, 0.2cm, 0.1cm}, clip]{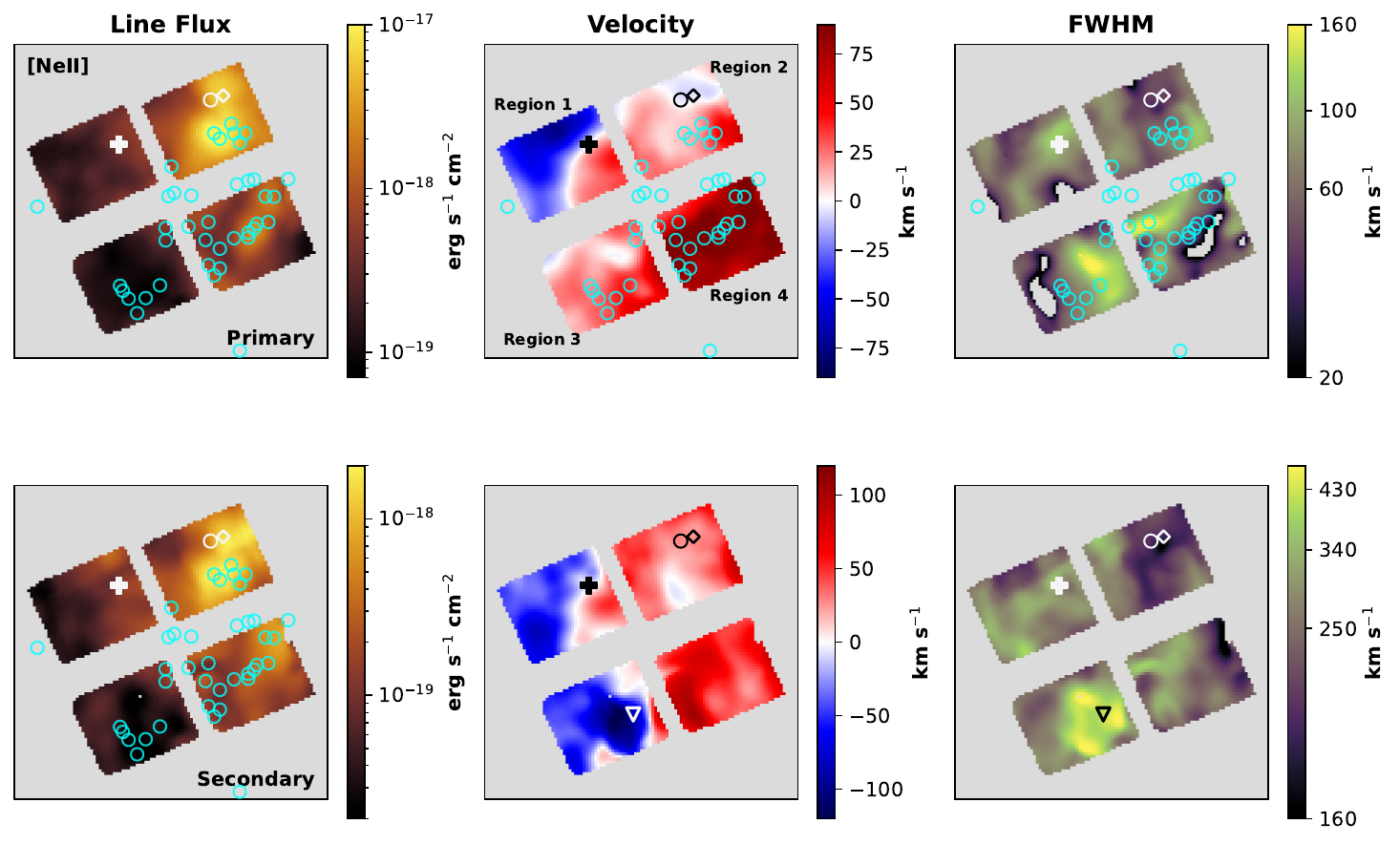}
\caption{
Flux, velocity shift, and FWHM of the \neii~12.8\micron{} line, with the narrow/primary component in the top row and the broader/secondary component in the bottom. Cyan circles again mark the position of UV-bright star clusters, and other symbols are as defined in Figure \ref{fig:h2slopes}.  
\looseness = -2
}
\label{fig:neon_maps}
\end{figure*}

The observed and attenuation-corrected fluxes of the S(1) -- S(7) transitions in the integrated spectrum are consistent within 5 - 10\% of the sum of the corresponding line fluxes from the spaxel-by-spaxel fits, with the greatest discrepancies occurring for the de-attenuated S(1) and S(3) fluxes. We attribute this to the modest attenuation ($\tau_{warm}$ = 0.49) inferred by {\tt CAFE} in our integrated spectrum, which is slightly lower than the median value of $\tau_{warm}$ = 0.60 for our resolved analysis as shown in Figure \ref{fig:cafetau}. 

\begin{table}[tb]
    \caption{{\sc~Model \molh{} Masses for M83's Nuclear Region}}
    \vspace{-0.3cm}
    \label{tab:masslist}
    \begin{center}
    \begin{tabular}[t]{c|ccc}
        \hline
        \hline
        Analysis & \tlprime{} & $M_{H2}(T >$ \tlprime) & $F_{CO}$ \\
         &  (K) & ($M_{\odot}$) & (\%) \\
        \hline
        Resolved & Model $T_l$ & $1.50^{+0.09}_{-0.09} \times 10^5$ & $0.89^{+0.03}_{-0.07}$\\
        & 200 & $2.26^{+0.17}_{-0.19} \times 10^5$ & $1.32^{+0.10}_{-0.11}$\\
        & 80 & $7.93^{+3.52}_{-2.59} \times 10^6$ & $46.2^{+20.5}_{-15.1}$\\
        & 50 & $8.17^{+8.67}_{-4.04} \times 10^7$ & $476^{+506}_{-235}$\\
        \hline
        Integrated & Model $T_l$ & $1.54^{+0.06}_{-0.07} \times 10^5$ & $0.90^{+0.04}_{-0.04}$\\
        & 200 & $1.84^{+0.09}_{-0.09} \times 10^5$ & $1.07^{+0.05}_{-0.05}$\\
        & 80 & $4.47^{+0.98}_{-0.84} \times 10^6$  & $26.1^{+5.72}_{-4.91}$\\
        & 50 & $3.05^{+1.14}_{-0.86} \times 10^7$ & $178^{+66.7}_{-49.9}$\\
        \hline
    \end{tabular}
    \tablecomments{\molh{} masses and uncertainties above a given temperature for both our resolved and integrated-light analyses. 
    Column 4 ($F_{CO}$) is the percentage of the CO-derived total molecular gas mass each measurement accounts for; we calculate $M_{H2}(CO)$ = 1.715 $\times 10^7 M_{\odot}$ assuming a Galactic $\alpha_{\mathrm{CO}}$ = 3.2 $M_{\odot}$ (K km s$^{-1}$ pc$^2$)$^{-1}$ and the CO(1-0) data of \citet{Hirota18}.}
    \end{center}
\end{table}

Additionally, while the best-fit value of $T_l$ (215.2 $\pm$ 26.2 K) is roughly consistent with the median $T_l$ value from the spaxel-level analysis, the power-law slope $n$ = 5.10 $\pm$ 0.26 is slightly lower (i.e., shallower) than the median slope in our resolved analysis, though still consistent within the uncertainties. Again adopting each of our three definitions of \tlprime, the \molh{} mass we infer from the integrated spectrum are (1) $1.54^{+0.06}_{-0.07} \times 10^5 M_{\odot}$ for \tlprime{} = 215.2 K; (2) $1.84^{+0.09}_{-0.09} \times 10^5 M_{\odot}$ for \tlprime{} = 200 K; and (3) $4.47^{+0.98}_{-0.84} \times 10^6 M_{\odot}$ for \tlprime{} = 80 K.

When adopting definition (1), the warm \molh{} mass estimates in our integrated and resolve-then-sum analyses are in excellent agreement with one another, as expected. However, in the temperature regimes considered under definitions (2) and (3), the \molh{} masses estimated from our integrated spectrum are at best only marginally consistent within uncertainties with those from our resolved analysis. The systematically lower masses in the integrated case can be attributed to the slight differences in typical power-law slope, which can greatly affect the inferred amount of cool/cold gas as we extend to lower and lower values of \tlprime. Together with the different degrees of attenuation inferred for integrated and spaxel-level data, this hints at a scenario where the inferred \molh{} mass and excitation depends on the physical scale probed, at least within the framework of the \ts{} model. We summarize the various estimates of \mht{\tlprime} and their uncertainties in Table \ref{tab:masslist}.

\subsection{Star Formation Rates}
\label{subsec:analysis:sfr}

As described in Section \ref{subsec:methods:emlines}, we performed single and double Gaussian fits to the \neii{} 12.81 \micron{} and \neiii{} 15.56 \micron{} lines. To determine whether to use the single or two-component fits for a given line in each spectrum, we compare the reduced-$\chi^2$ of the two fits in a statistical $F$-test, following \citet{Westmoquette07}. In essence, we adopt the two-component fit if the reduced-$\chi^2$ is lowered by a factor of at least 3.289 compared to the one-component fit, corresponding to a significance level $\alpha = 0.10$; otherwise, we default to the one-component fit. Only after application of the $F$-test do we then apply our detection threshold of S/N $>$ 3, since our definition of noise depends on the fitted FWHM of the line or line component.

\begin{figure}[tb]
\centering
\includegraphics[width=0.99\linewidth,trim={0.2cm, 0.1cm, 0.2cm, 0.1cm}, clip]{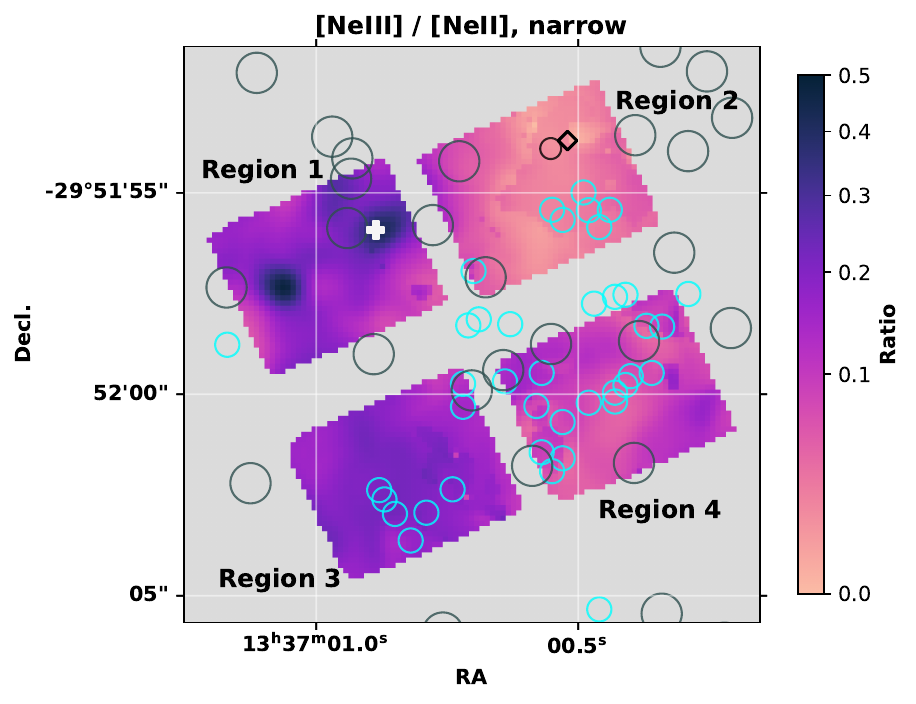}\\
\includegraphics[width=\linewidth,trim={0.2cm, 0.1cm, 0.2cm, 0.1cm}, clip]{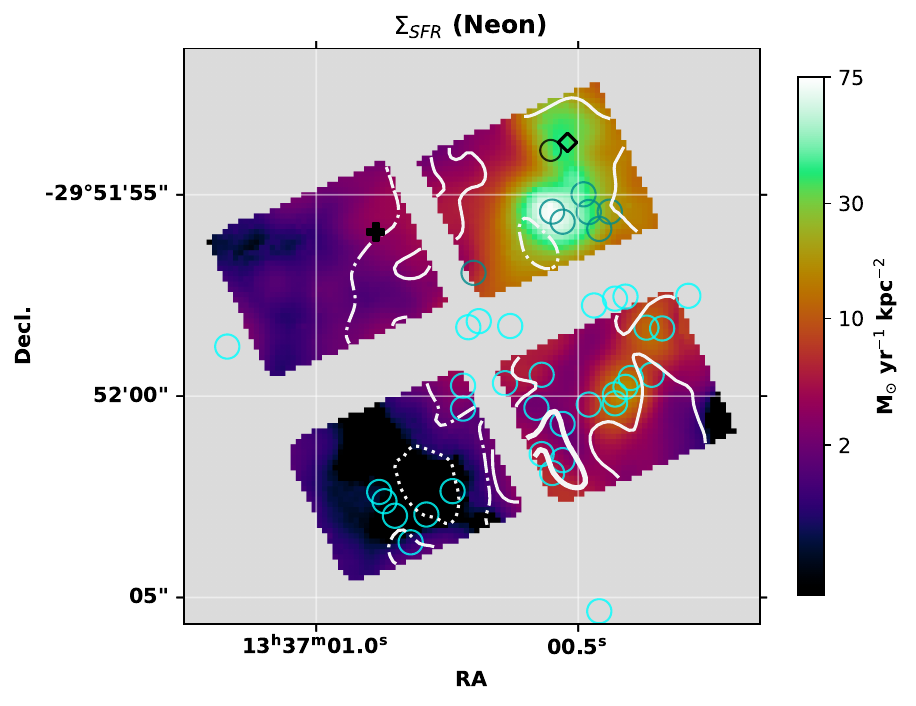}
\caption{
{\em Top:} Line flux ratio of \neiii/\neii, considering only the narrow component of \neii{} and \neiii{} (or the single Gaussian fit, if a two-component fit was not needed for some spectra).
{\em Bottom:} SFR surface density across our MRS cube as computed using the diagnostic of \citet{Zhuang19}. Cyan circles mark the position of UV-bright star clusters from \citet{Harris01}, and grey circles mark the locations of supernova remnants from \citet{Winkler23}. White contours in the lower panel show curves of constant velocity $v \in$ [-90, -40, 0, 40, 90] km s$^{-1}$ for the secondary component of \neii.
\looseness = -2
}
\label{fig:neonSFR}
\end{figure}

The bright \neii{} line is better modeled with a two-component fit across nearly all of our FOV (see Figure \ref{fig:neon_maps}), indicating a ubiquitous ISM component that is kinematically distinct from gas near H{\sc ii} regions. The secondary component contributes $\sim$23\% of the total \neii{} line flux on average in our FOV. The need for a second component in \neiii{} is less uniformly evident, with our $F$-test favoring a single Gaussian fit across large swathes of all four regions. 
However, we note that for \neiii{} spectra which do not ``pass" the $F$-test, the FWHM of the single-Gaussian fits are slightly but systematically larger (by 20 -- 60 km s$^{-1}$) than the FWHM of the primary component in the corresponding double-Gaussian fit. This suggests that there may be a broad component that is strong enough to affect the wings of the \neiii{} line profile, but is unable to be detected by the $F$-test at a reasonable statistical significance. We show the results of both our single-Gaussian and double-Gaussian fitting to \neiii{} in Appendix \ref{sec:appendix} (Figure \ref{fig:appendix:neon3}).

We additionally show the flux ratio of the primary/narrow components of \neiii{} to \neii{} in the upper panel of Figure \ref{fig:neonSFR}. We find that the neon emission is predominantly from the singly-ionized state across our FOV with the exception of two compact regions in Region 1, where \neiii{} and \neii{} are of roughly equal strength. One of these areas corresponds to the location of the optical nucleus, while the other has an unclear origin. Though shocks may be partly responsible for the elevated \neiii{} abundance, the nearest catalogued supernova remnant \citep[][SNR, marked with grey circles in Figure \ref{fig:neonSFR}]{Winkler23} is 1\farcs4 (33 pc projected) away.

Finally, we use our $F$-test and S/N-filtered line luminosity maps to derive SFRs per spaxel using the relation presented in \citet{Zhuang19}. Their relation is similar to that given in \citet{Ho07} but with an explicit dependence on metallicity, which affects the neon abundance and the shape of the incident ionizing spectrum. \citet{Zhuang19} also provide prescriptions for determining the fraction of neon in the first and second ionization states, $f_+$ and $f_{+2}$, directly from the \neiii/\neii{} line ratio. We use the extinction-corrected line luminosities and an assumed gas-phase metallicity of $Z = 2\times Z_{\odot}$ \citep{Hernandez21} to derive SFRs using the \citet{Zhuang19} relation. Converting the spaxel size (0\farcs13) to kpc at the distance of M83 \citep[$D$ = 4.6 Mpc][]{Saha06}, we create a map of the SFR surface density $\Sigma_{SFR}$ in $M_{\odot}$ yr$^{-1}$ kpc$^{-2}$, shown in the lower panel of Figure \ref{fig:neonSFR}. The SFR densities in Region 2 are fairly high, reaching $\Sigma_{SFR}$ $\sim$75 $M_{\odot}$ yr$^{-1}$ kpc$^{-2}$ near a grouping of very young (see Section \ref{subsec:disc:stararc}) UV-bright star clusters, and rapidly decrease as the eye moves clockwise through Regions 4, 3, and 1, with SFR densities as low as $\sim$1 $M_{\odot}$ yr$^{-1}$ kpc$^{-2}$.

We note the \citet{Zhuang19} prescriptions were calibrated using Starburst99 SSP models \citep{starburst99} only up to ages of 2.5 Myr, while the stellar populations in our field span ages up to $\gtrsim$10 Myr.
The star cluster ages ($\lesssim$3 Myr) we estimate from archival {\em HST}/COS spectroscopy in Regions 2 and 4 (see Section \ref{subsec:disc:stararc}) are roughly consistent with this calibration. Slightly older stellar populations are mainly seen in Region 3 (e.g., 8.6 Myr in the photometric analysis from \citet{Knapen10}). This area is where we see the faintest \neii{} and \neiii{} emission, consistent with a low instantaneous SFR despite being outside the age calibration of the \citet{Zhuang19} relations.

\section{Discussion}
\label{sec:disc}

\subsection{Molecular Gas Contents of M83's Nuclear Region}
\label{subsec:disc:molgas}

Historically, attempts to quantify the amount and distribution of molecular gas in M83 \citep{Hirota18, Koda23} and SF galaxies more generally \citep{Bolatto13, Pavesi18, Leroy21} have relied on observations of CO (and other, less abundant tracers -- e.g., OH; \citealt{Xu16}) which, unlike \molh, have a nonzero dipole moment and hence are strong rotational emitters. CO-based molecular gas measurements usually require an assumed value for the  factor $\alpha_{\mathrm{CO}}$ to convert between the intensity of CO emission and \molh{} mass surface density, although the value of $\alpha_{\mathrm{CO}}$ can vary significantly with time, environment, and metallicity, even within individual galaxies \citep{Sandstrom13, Hu22, Chiang24, Lee24}.

Our use of the \ts{} model, in principle, allows us to extract estimates of the total molecular gas mass across our FOV in addition to only the warm ($T > 200$ K) \molh. \ts{} distinguishes between the lower temperature bound for the power-law model $T_l$ and the sensitivity temperature $T_s$, or the temperature below which an increase in the \molh{} mass results in no measurable change to the line fluxes or excitation, which they determine following \citet{Avni76} by evaluating their model over a grid of $n$ and $T_l$ values and examining the change in $\chi^2$ relative to its minimum value, $\Delta\chi^2$. For the {\em Spitzer}/IRS spectra they considered in that work, only a handful of galaxies (primarily LINER and Seyfert hosts) showed evidence for a well-defined minimum in $\Delta\chi^2$, such that extrapolating the power-law model for those galaxies to temperatures less than $\sim$120 K results in a poorer fit. They interpret this as evidence for a warm gas excess in these sources, such that the majority of \molh{} is directly detected via the mid-IR rotational transitions. Most galaxies in their sample have no well-defined minimum in $\Delta\chi^2$ space, meaning that any $T_l < T_s$ remains consistent with the observed line ratios; their sample average is $T_s$ = 81 K.

For completeness, we performed the same $\Delta\chi^2$ evaluation exercise as in \ts{} for select spaxel-level spectra and for the FOV-integrated spectrum to obtain estimates of the sensitivity temperature for our observations. We find in all cases that there is a well-defined minimum in $\Delta\chi^2$, such that extrapolating below the best-fit lower temperature inferred by the model ($T_l \sim$220 -- 270) degrades the fit. However, we stress that unlike in \ts, our data and analysis does not include the low-excitation S(0) line. Without this line to constrain the colder \molh, the power-law temperature model has little incentive to ``prefer" low temperatures at a fixed slope $n$. We therefore cannot directly constrain a sensitivity temperature comparable to that in \ts{} for the observations presented here. However, as discussed in Sec.\ \ref{subsec:analysis:h2mass}, we nonetheless estimated the cool+warm \molh{} mass down to 80 K, $7.93^{+3.52}_{-2.59} \times 10^6 M_{\odot}$, to serve as a lower limit to the total amount of molecular gas in M83's nuclear region.

\ts{} also discusses the utility of a calibration temperature, which we label here as $T_{cal}$, defined such that at $T_l$ = $T_{cal}$ the \molh{} mass inferred by their model equals the cold/total molecular gas mass derived from CO emission, \mht{$T_{cal}$} = $M_{H2}$(CO). In that work they assume a Galactic $\alpha_{\mathrm{CO}}$ = 3.2 $M_{\odot}$ (K km s$^{-1}$ pc$^2$)$^{-1}$, which we also adopted in H23 and the present work. \ts{} report an average $T_{cal}$ = $49 \pm 9$ K for their {\em Spitzer}/IRS sample, such that \molh{} warmer than $\sim$50 K can account for the entirety of the molecular gas mass inferred from CO in their galaxies.

We performed a first-pass analysis of this type in H23, in which we applied the \ts{} model to the integrated spectra of our four regions. In that work, we measured a warm \molh{} mass of $8.76 \pm 0.33 \times10^4 M_{\odot}$ summed across the four regional spectra, tracing \molh{} from 2000 K down to 252 -- 301 K. We also measured $M_{H2}(T > 50$ K) = $6.79 \pm 0.54 \times 10^7 M_{\odot}$, which we took as an estimate of the total \molh{} mass in our field.

Recalling the three definitions of \tlprime{} we outlined in Section \ref{subsec:analysis:h2mass}, the warm \molh{} mass estimate from H23 is most consistent with definition (1), which uses the best-fit value of $T_l$ as the lower bound on the mass integral. Using that limit, in the present work we measured a warm \molh{} mass of $1.50^{+0.09}_{-0.09} \times 10^5 M_{\odot}$ summed across our FOV. Although the nominal warm gas mass we measure in this work is higher by a factor of $\sim$2, we note that H23 did not apply attenuation corrections to their line fluxes. Since the \molh{} mass estimate is proportional to the S(1) line flux in the \ts{} framework, the difference between the two values is almost entirely accounted for by our use of attenuation corrections in the present work.

If we instead extrapolate the best-fit \ts{} models for each pixel to $T_{cal}$ = 50 K, we obtain $M_{H2}(T > 50$ K) = $8.17^{+8.67}_{-4.04} \times 10^7 M_{\odot}$. Although the nominal ``total" mass is again slightly higher than that reported in H23, it remains consistent within the (considerable) uncertainties, especially once attenuation corrections are taken into account. The bulk of the uncertainty on our new estimate of the total molecular gas mass is attributable to uncertainties on the per-pixel power-law slopes.

\subsection{Potential CO-dark Molecular Gas}
\label{subsec:disc:codark}

The observational dependence on CO introduces the problem of ``CO-dark" molecular gas \citep[e.g.,][]{Wolfire10}, or reservoirs of \molh that are not traced by cospatial CO emission. As FUV photons and cosmic rays \citep[CRs; e.g.,][]{Bisbas17} permeate into a molecular cloud, the CO molecules dissociate and leave behind an envelope of ionized and neutral carbon in addition to the colder, denser cloud core. \molh, meanwhile, can self-shield from photodissociation \citep{Gnedin14} by FUV photons, which leaves a potentially significant reservoir of gas that is coinhabited (partially or fully) by \molh, C, and C$^+$ but not CO \citep{Glover12, Madden20, Hu21}.

In H23, we used the \ts{} framework to estimate the \molh{} gas mass \mht{50 K} in the integrated spectra of our four regions, which we took as a measure of the total molecular gas mass. Using the CO(1-0) observations of \citet{Hirota18} and a Galactic CO-to-\molh{} conversion factor ($\alpha_{\mathrm{CO}}$ = 3.2 $M_{\odot}$ (K km s$^{-1}$ pc$^2$)$^{-1}$), we compared this ``total" molecular gas mass per region to that inferred from CO using traditional methods and found \mht{50 K} $>$ $M_{H2}$(CO) in all four pointings. We estimated that, depending on which region/environment we consider, between $\sim$27\% and 92\% of the molecular gas mass is not accounted for through CO(1-0) emission -- in other words, CO-dark gas mass fractions upwards of $\sim$30\%, even in the super-solar metallicity core of M83.

While a direct comparison of pixel-level gas masses between our MRS observations and CO data from ALMA is beyond the scope of this work, the resolved power-law models presented above allow us to more accurately examine the CO-dark gas contents of M83's nuclear region when considering the entirety of our FOV. As summarized in Table \ref{tab:masslist}, at temperatures $>$80 K, the \ts{} model can only recover $\sim$46\% of the CO-based molecular gas mass in our field. However, given the moderately steep power-law slopes we measure in some areas (Figure \ref{fig:h2slopes}), extrapolating to lower temperatures will rapidly inflate the predicted molecular gas mass. At \tlprime{} = 50 K, the total $M_{H2}$ mass is a factor of $\sim$4.7 higher than that inferred from CO observations, implying that $\sim$79\% of the molecular gas content of M83's nuclear region is not traced by CO.

On the other hand, if we assume that the CO-based gas mass estimates are correct, we may tweak the chosen value of \tlprime{} to find an appropriate $T_{cal}$ for our observations, such that \mht{$T_{cal}$} = $M_{H2}$(CO). We iterated over values of \tlprime{} to find a range of temperatures that satisfy \mht{\tlprime} = $M_{H2}$(CO), again only considering the sum of our FOV. For our resolved analysis, a range of temperatures $T_{cal}$ (60 -- 70 K) yield an \molh{} mass estimate that is consistent with $M_{H2}$(CO) = 1.715$\times 10^7 M_{\odot}$ within the uncertainties. For the integrated-light spectrum, we find that $T_{cal}$ = 57 K provides the best match to $M_{H2}$(CO). These are generally consistent with the range of $T_{cal}$ values given in Table 4 of \ts{} for their sample of SINGS galaxies.

While temperatures $\gtrsim$50 K are significantly higher than the $\sim$10 -- 30 K range typically assumed for the dense, imminently star-forming parts of molecular clouds, the prediction of significant CO-dark gas reservoirs even in high-metallicity galaxies has been borne out repeatedly in simulations. For example, in the simulations by \citet{Hu21}, their models with time-dependent \molh{} chemistry suggest a shallow radial abundance profile for the \molh{} gas such that a significant fraction of the \molh{} mass is found in low-density regions, far from the CO-bearing core. When showing the density distribution in their simulations weighted by CO mass, \citet{Hu21} find that the bulk of the CO is found at densities comparable to or greater than the densities at which C is readily converted to CO. When weighting by \molh{} mass, however, the majority of \molh{} is found at densities significantly lower than the H/\molh{} conversion density. This suggests that a significant fraction of the \molh{} mass is found not in the CO-bearing cloud cores, but in the atomic-dominated regime. While this effect is most extreme at low metallicities (and hence intrinsically low CO abundances), even at $Z = 1 - 3 \times Z_{\odot}$, between 20 and 35\% of the \molh{} mass can be found in gas with abundances $log_{10}$(2$x_{H_2}$) $<$ -0.5 in the \citet{HU21} simulations. When translated to a mass fraction of CO-dark gas at (super-)solar metallicities, they report that even high-sensitivity CO(1-0) observations (down to column density limits of $\sim$6 $\times 10^{14}$ cm $^{-2}$), $\sim$55 -- 60\% ($\sim$58 -- 70\%) of the \molh{} gas mass is CO-dark in their time-dependent (steady state) models. 

Even in the context of \citet{Hu21} and other theoretical works which predict the existence of CO-dark gas in galaxies \citep{Wolfire10, Smith14}, CO-dark mass fractions of $\sim$80\% in the $\gtrsim2\times Z_{\odot}$ environments of M83's nucleus are surprising and reinforce the need for small-scale spectroscopic observations of the ISM in starbursting galaxies at all metallicities. It is worth noting that our estimate of $M_{H2}$(CO) in our FOV already has a CO-to-\molh{} conversion factor baked in, which (in principle) should account for the fact that CO does not perfectly trace the low-density \molh. The fact that our \molh{} mass measurements at cool temperatures $\gtrsim70$ K (let alone the colder, denser ISM) are comparable to $M_{H2}$(CO) suggests that a higher-than-Galactic value of $\alpha_{\mathrm{CO}}$ may be necessary even in solar or supersolar environments. A possible exception to this is the nucleus proper, which has the highest best-fit values of $T_l$ ($>$300 K) in the \ts{} framework (Figure \ref{fig:h2slopes}). This suggests that the bulk of the \molh{} in the central $r \sim 15$ pc of the galaxy is in a warm excess phase and is detected directly from rotational emission.

The notion that high values of $\alpha_{\mathrm{CO}}$ are needed for the nuclear region of M83 runs contrary to several literature results which suggest galaxy centers, starbursting galaxies, and galaxies with high metallicity are expected to have $\alpha_{\mathrm{CO}}$ factors lower than the Galactic value (e.g., $\alpha_{\mathrm{CO}} \lesssim 1.0~M_{\odot}$ (K km s$^{-1}$ pc$^2$)$^{-1}$, vs.\ Galactic values $\alpha_{\mathrm{CO}} \approx 3 - 4~M_{\odot}$ (K km s$^{-1}$ pc$^2$)$^{-1}$); see \citet{Bolatto13} and references therein. For example, recent work by \citet{Lee24} shows that, within M83's disk, $\alpha_{\mathrm{CO}}$ decreases steadily with radius from $\sim 5 - 7 M_{\odot}$ (K km s$^{-1}$ pc$^2$)$^{-1}$) at $R = 6$ kpc to $\sim 2 M_{\odot}$ (K km s$^{-1}$ pc$^2$)$^{-1}$) at $R = 2$ kpc. ALMA observations of barred galaxy centers at $\sim$100 pc scales \citep{Teng22, Teng23} find generally low values of $\alpha_{\mathrm{CO}}$ ($\lesssim 0.5 M_{\odot}$ (K km s$^{-1}$ pc$^2$)$^{-1}$) that decrease with galactocentric radius, driven primarily by a correlation with the optical depth of CO emission.

In the context of these results, a low assumed value of $\alpha_{\mathrm{CO}}$ may be more appropriate for M83's nuclear region than the value we adopted in this work (3.2 $M_{\odot}$ (K km s$^{-1}$ pc$^2$)$^{-1}$). However, a low $\alpha_{\mathrm{CO}}$ would result in even lower estimates of $M_{H2}$(CO) in our field, exacerbating the discrepancy with the \molh{} masses from the \ts{} model which include the coldest ($\lesssim$60 K) gas. Comprehensive modeling of the \molh{} rotational lines, CO lines, dust emission, and other putative tracers of CO-dark gas \citep[e.g., the $^3P_1 \rightarrow ^3P_0$ of C{\sc i}][]{Bisbas17, Papado22, Ramambason24} will help to better understand the relationship between diffuse \molh{} and traditional molecular gas tracers in heavily star-forming environments.

\begin{figure}[t]
\centering
\includegraphics[width=0.75\linewidth,height = 1.8\linewidth,trim={0.2cm, 0.1cm, 0.2cm, 0.1cm}, clip]{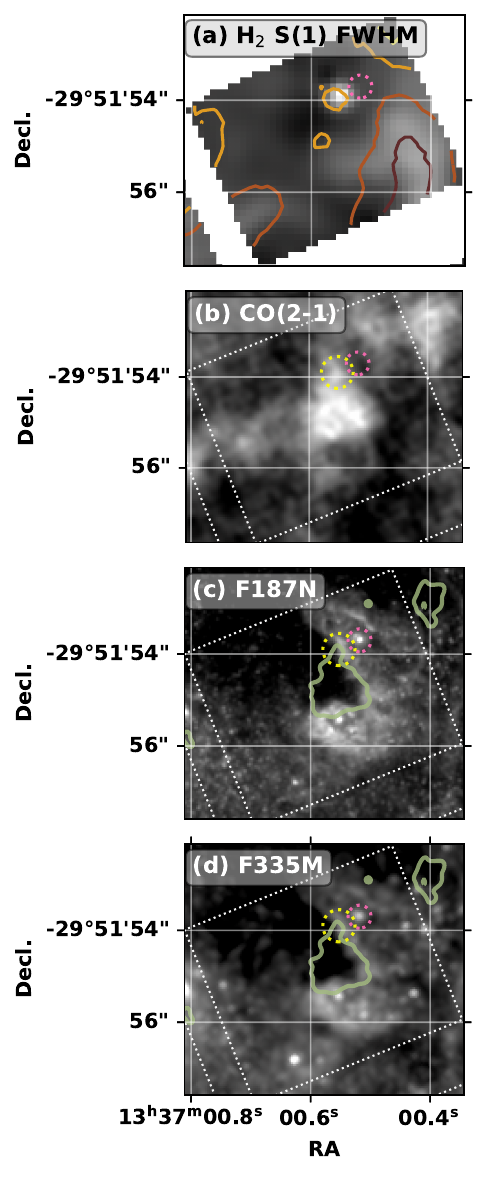}\\
\caption{Zoom-in on Region 2, highlighting the multi-band properties of an unknown, obscured point source. From top to bottom, we show (a) the \molh{} S(1) FWHM map; (b) Single-channel CO(2-1) emission at $-20$ km s$^{-1}$; (c) Pa$\alpha$ imaging from JWST-FEAST; and (d) F335M emission from JWST-FEAST.
Brown-yellow contours in (a) show the \molh{} S(1) flux, while the green contour in (c) and (d) mark regions with CO(2-1) intensity $>150$ mJy beam$^{-1}$. The obscured point source and a nearby reddened star cluster are respectively marked with yellow and pink dotted circles.
\looseness = -2
}
\label{fig:babycluster_zoom}
\end{figure}

\subsection{A Massive, Obscured Point Source}
\label{subsec:disc:baby}

We now briefly discuss a point source which we identify through the flux and kinematic maps of the S(1) line shown in the top row of Figure \ref{fig:h2vis}. Specifically, we identify a compact ($\lesssim$5 pc) source at ($\alpha$, $\delta$) = (13:37:00.555, -29:51:53.900) which (1) is slightly redshifted (by $\sim$10 km s$^{-1}$) relative to its surroundings and (2) is fairly broad and spectrally resolved, with a FWHM of $>$150 km s$^{-1}$ at its core. Additionally, this kinematic structure appears at high significance {\em only} in the S(1) line, with neither Ne lines nor other \molh{} rotational lines indicating the presence of a point source at the same position. 

We also note a moderately reddened (possibly partially-embedded) star cluster at ($\alpha$, $\delta$) = (13:37:00.5205, -29:51:53.702), as measured from {\em JWST}/NIRCam imaging from the FEAST program (see the next section). The coordinates of the cluster (marked with a diamond in Figures \ref{fig:h2slopes} and \ref{fig:nucmulti}) and of the kinematic hot spot differ by $\sim$0\farcs43 or 9.6~pc.
We show multi-wavelength zoom-ins on this area in Figure \ref{fig:babycluster_zoom}, including the FWHM map of the \molh{} S(1) line; CO(2-1) emission (ALMA pID 2013.1.01161.S, PI Sakamoto), and F187N and F335M imaging from JWST-FEAST (Adamo et al.\ in prep.). The F187N image in Fig.~\ref{fig:babycluster_zoom}(c) clearly shows that the partially-embedded cluster has blown away most of its natal material. The ionized shell is a quite striking feature that is `pinched' in the NW and SE direction, presumably due to higher density material, and more freely expanding to the NE. The `pinch' point to the SW corresponds to the edge of a CO cloud (Fig.~\ref{fig:babycluster_zoom}(b)) and the location of the \molh{} point source.

Recent studies of the spiral galaxy NGC 1365 from the PHANGS collaboration \citep{Whitmore23} showed that YMCs with ages $\sim 1.3 \pm 0.7$ Myr are fully obscured in the optical and can remain completely or partly obscured for up to $\sim 3.7 \pm 1.1$ Myr. This is consistent with clearing timescales of 3 -- 5 Myr based on {\em HST} data only, as determined for M83 \citep{Hollyhead15, Deshmukh24} and other galaxies \citep[e.g.,][]{Messa21, McQuaid24}.
The stellar population modelling we discuss in Section \ref{subsec:disc:stararc} suggests very young ages (2 $\pm$ 1 Myr) even for the UV-brightest and largely unobscured clusters in our Region 2. The young stellar ages in this area, along with the spatial coincidence of the \molh{} source with CO-emitting gas, leads us to speculate that the kinematically-distinct \molh{} point source may be a forming or fully-embedded star cluster.

\subsection{Gaseous Confirmation of a Stellar Age Gradient?}
\label{subsec:disc:stararc}

Nearby starbursts, such as the nuclear region in M83, make optimal laboratories for unraveling the mechanisms operating in extreme environments and across different phases of star formation as their proximity affords exquisite spatial resolution. This is clearly highlighted in the true-color {\em HST} optical image displayed in Figure \ref{fig:interestingplaces}(e). This figure shows the various complexities in this starburst system, with some regions almost completely obscured towards the north (in Region 2), partially cleared ISM channels in the west-southwest area (Region 4), and fully cleared regions towards the southeast (Region 3).

\begin{figure*}[t]
\centering
\includegraphics[width=6.5cm,height=5.5cm,trim={0.2cm, 0.1cm, 0.2cm, 0.1cm}, clip]{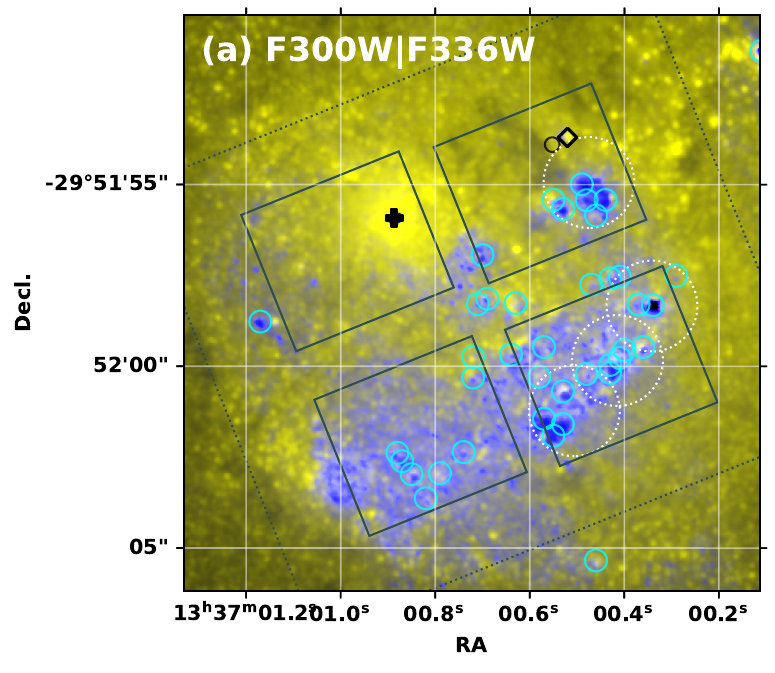}
\includegraphics[width=5.0cm,height=5.5cm,trim={2.9cm, 0.1cm, 0.2cm, 0.1cm}, clip]{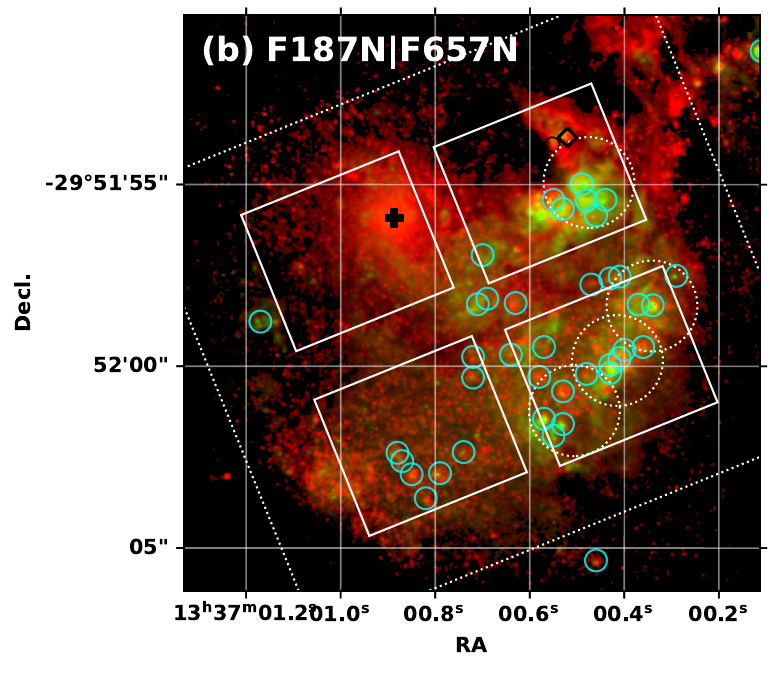}
\includegraphics[width=5.0cm,height=5.5cm,trim={2.9cm, 0.1cm, 0.2cm, 0.1cm}, clip]{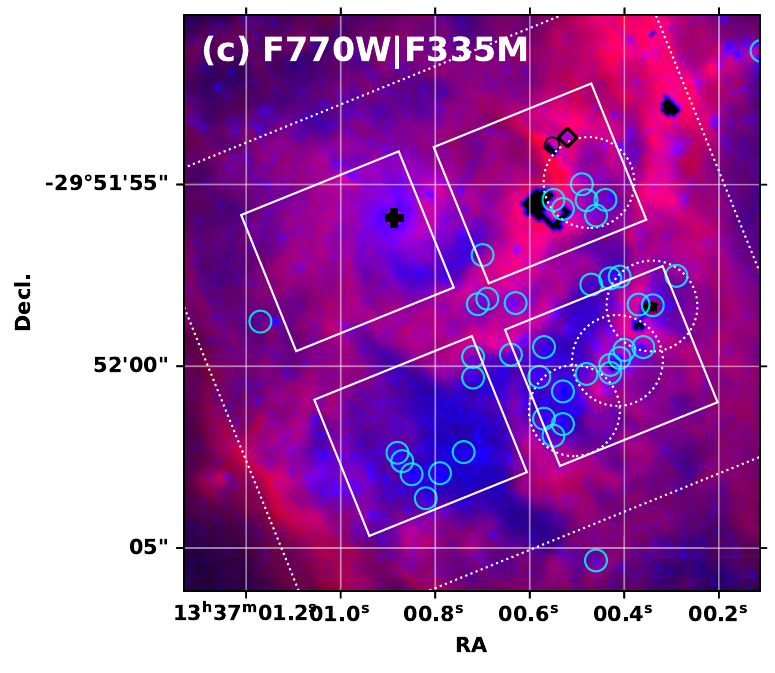}\\
\includegraphics[width=6.5cm,height=5.5cm,trim={0.2cm, 0.1cm, 0.2cm, 0.1cm}, clip]{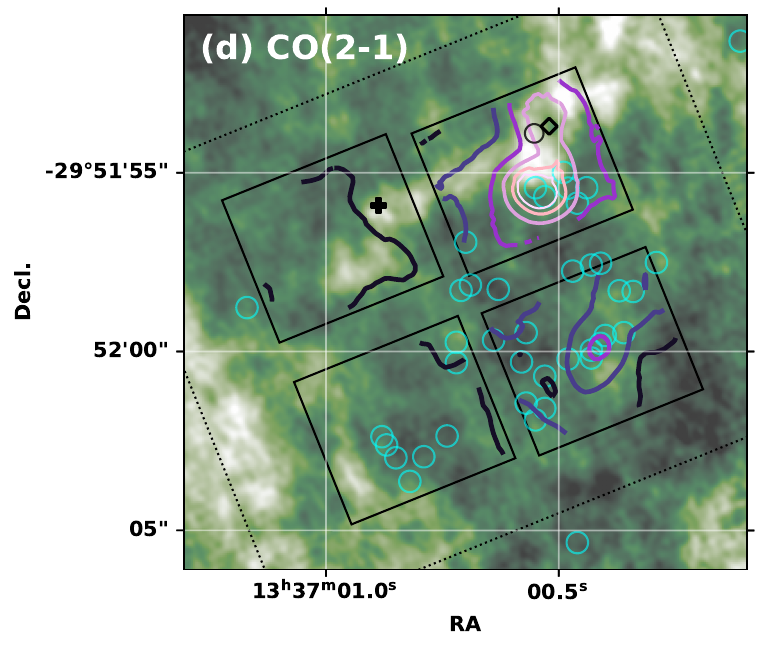}
\includegraphics[width=5.0cm,height=5.5cm,trim={2.9cm, 0.1cm, 0.2cm, 0.1cm}, clip]{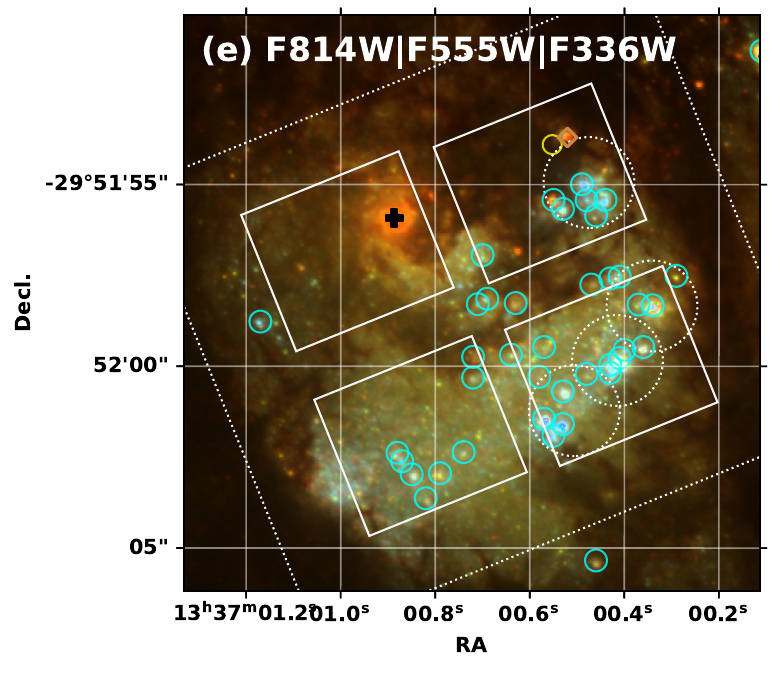}
\includegraphics[width=5.0cm,height=5.5cm,trim={2.9cm, 0.1cm, 0.2cm, 0.1cm}, clip]{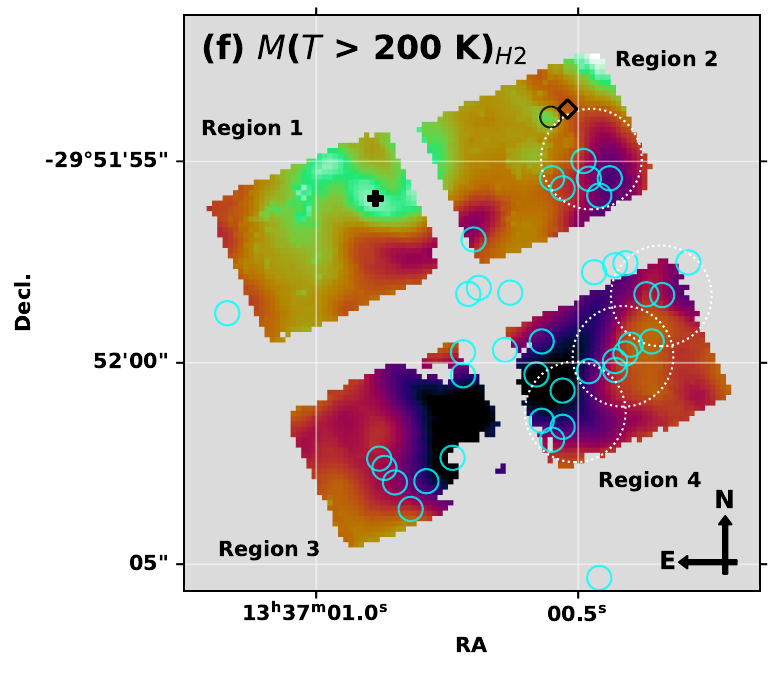}
\caption{Multiwavelength view of the center of M83 using images from {\em HST}/WFC3, {\em JWST}/NIRCam, {\em JWST}/MIRIM, and ALMA.
The brightness of some images have been rescaled to better show small-scale structures. Solid squares denote the Ch1 FOV of our four MRS pointings, and the dotted square approximates the total Ch4 FOV. White-dotted and cyan circles are as described in previous figures. The plus symbol, diamond, and variably-colored circle respectively note the position of the optical nucleus, a partially-embedded YMC, and an unidentified \molh{} point source described in Section \ref{subsec:disc:baby}. 
(a) Old (yellow) and young (blue) stellar populations as traced by NIRCam F300W (JWST-FEAST) and WFC3 F336W (GO 14059, PI Soria).
(b) Differentially extincted ionized gas as traced by Pa$\alpha$ (red; JWST-FEAST) and H$\alpha$ (green; GO 11360, PI O'Connell). 
(c) Ionized PAH emission from the 7.7 (red) and 3.3 \micron{} (blue) features (JWST-FEAST).
(d) CO(2-1) data from ALMA (2022.1.00951, PI Thater). Purple-pink contours trace $\Sigma_{SFR}$ using the \citet{Zhuang19} diagnostic at intervals of [2, 5, 15, 30, 45, 55] $M_{\odot}$ yr$^{-1}$ kpc$^{-2}$.
(e) True-color optical image using WFC3 F814W, F555W, and F336W (GO 14059, PI Soria).
(f) Warm \molh{} mass distribution (at temperatures $>$ 200 K) from this work.
\looseness = -2
}
\label{fig:nucmulti}
\end{figure*}

The {\em JWST} and {\em HST} color image in panel (a) of Figure \ref{fig:nucmulti} shows the spatial distribution of the stellar populations, specifically contrasting the old stars (500--1000 Myr; shown in yellow) with young stars ($<$ 25 Myr; shown in blue). The younger, bluer population in Figure \ref{fig:nucmulti}(a) depict a clear ``active ringlet," as labeled by \citet{Harris01}, who find an apparent age gradient through a photometric study investigating the properties of the massive clusters in this circumnuclear region. Using {\em HST} broadband and narrowband filters, \citet{Harris01} report that stellar clusters with ages $<$ 5 Myr are preferentially located along the northwestern area (Region 2 in this work). The presence of this age gradient along the starburst's arc, running from northern to southeastern regions, has since then been studied and debated \citep{Diaz06, Knapen10, Wofford11}.

To further investigate the possible age gradient, we have analyzed archival {\em HST}/COS observations targeting YMCs in Regions 2 and 4. In the final two panels of Figure \ref{fig:interestingplaces}, we show with white-dotted circles the locations of the four COS pointings, which encompass multiple star clusters in a single 2\arcsec.5 aperture. We characterize the dominant stellar population in each COS pointing using the \texttt{SESAMME} software \citep{Jones2023}. Being a Python-based full spectrum fitting code, \texttt{SESAMME} uses Markov Chain Monte Carlo methods to obtain estimates of the cluster age, metallicity, E(B-V) values, and when the distance to the clusters is known, the user can also constrain the cluster mass. A caveat to the analysis of the COS observations with this software is that \texttt{SESAMME} assumes individual stellar populations for each input spectrum. In the context of our test, this means that our analysis then assumes a single dominant stellar population when inferring the stellar properties of the FUV integrated-light spectra. We find that the most massive stellar population encompassed in our COS pointings resides in Region 2, with a mass estimate of 1.13 ($\pm$ 0.02) $\times$ 10$^{6}$ M$_\odot$ and an age of 2 $\pm$ 1 Myr. For the remaining targets in Region 4 we infer an age of 3 $\pm$ 1 Myr for the two easternmost clusters, and 2 $\pm$ 1 Myr for the pointing west of Region 4. Generally, this agrees with the originally proposed trend in the active ringlet, where some of the youngest stellar clusters are found in Region 2. We note that the science programs collecting the {\em HST}/COS archival data ({\em HST} program IDs: 11579, 14681 and 15193) analyzed here primarily targeted the brightest FUV objects in the nuclear region of M83. Not surprisingly, we estimate very young ages for these clusters. \par

In the context of the ``active ringlet", the spaxel-level analysis we present here provides additional support for this age gradient.
For example, the bulk of star-formation activity in our FOV (with $\Sigma_{SFR} > 50$  $M_{\odot}$ yr$^{-1}$ kpc$^{-2}$) is found in a clump coincident with (or just eastward of) the massive grouping of very young, unobscured star clusters in Region 2 (Figure \ref{fig:neonSFR}), plus a plume of high SFR extending northward. Conversely, the lowest SFRs are generally found in Region 3, where the \neii{} emission is significantly broadened and blueshifted. 

Each of these structures -- the low-SFR patch and the high-SFR clump and plume -- also appear in other figures shown in this work. The low-SFR patch, for example, is coincident with a region of faint CO(2-1) and \molh{} emission and fairly steep power-law slopes in the \ts{} model, suggesting little molecular gas mass and low \molh{} excitation. This patch also shows low optical extinction and is coincident with slightly evolved stellar populations (8.6 Myr in region A of \citealt{Knapen10}). While we do not attempt to distinguish (and indeed cannot in the present analysis) a specific mechanism that drove the creation of this dust-less patch (e.g., photoionization vs.\ shocks), the combination of low-density ISM, ionizing outflows, and apparent lack of $<$5 Myr-old stellar populations in this area strongly suggests a link to previous bursts and cessations of star formation. 

The high-SFR clump and plume, meanwhile, roughly corresponds to a region with both high \molh{} excitation (Figure \ref{fig:h2slopes}) and modest warm \molh{} mass (Figure \ref{fig:nucmulti}(f)). Interestingly, the CO(2-1) emission is also fairly weak in parts of the clump, where we see bright, ringed structures in Pa$\alpha$, F335M, and F770W imaging from the JWST-FEAST program (Figures \ref{fig:babycluster_zoom}(c,d), \ref{fig:nucmulti}(c)), which suggests that recently-formed stars may be destroying CO and exciting \molh{} in this area. Conversely, the strongest CO(2-1) emission and highest warm \molh{} surface densities are more coincident with a knot of high obscuration which extends to the northeast of Region 2, including the kinematic source seen in \molh{} S(1). This knot is dark even in {\em JWST}/NIRCam imaging (Figure \ref{fig:nucmulti}(a,b,e)), suggesting extremely high column densities.

\subsection{Spatially-varying ISM Properties in M83}
\label{subsec:disc:interestingplaces}

\begin{figure*}[t]
\centering
\includegraphics[width=0.95\linewidth,trim={0.2cm, 0.1cm, 0.2cm, 0.1cm}, clip]{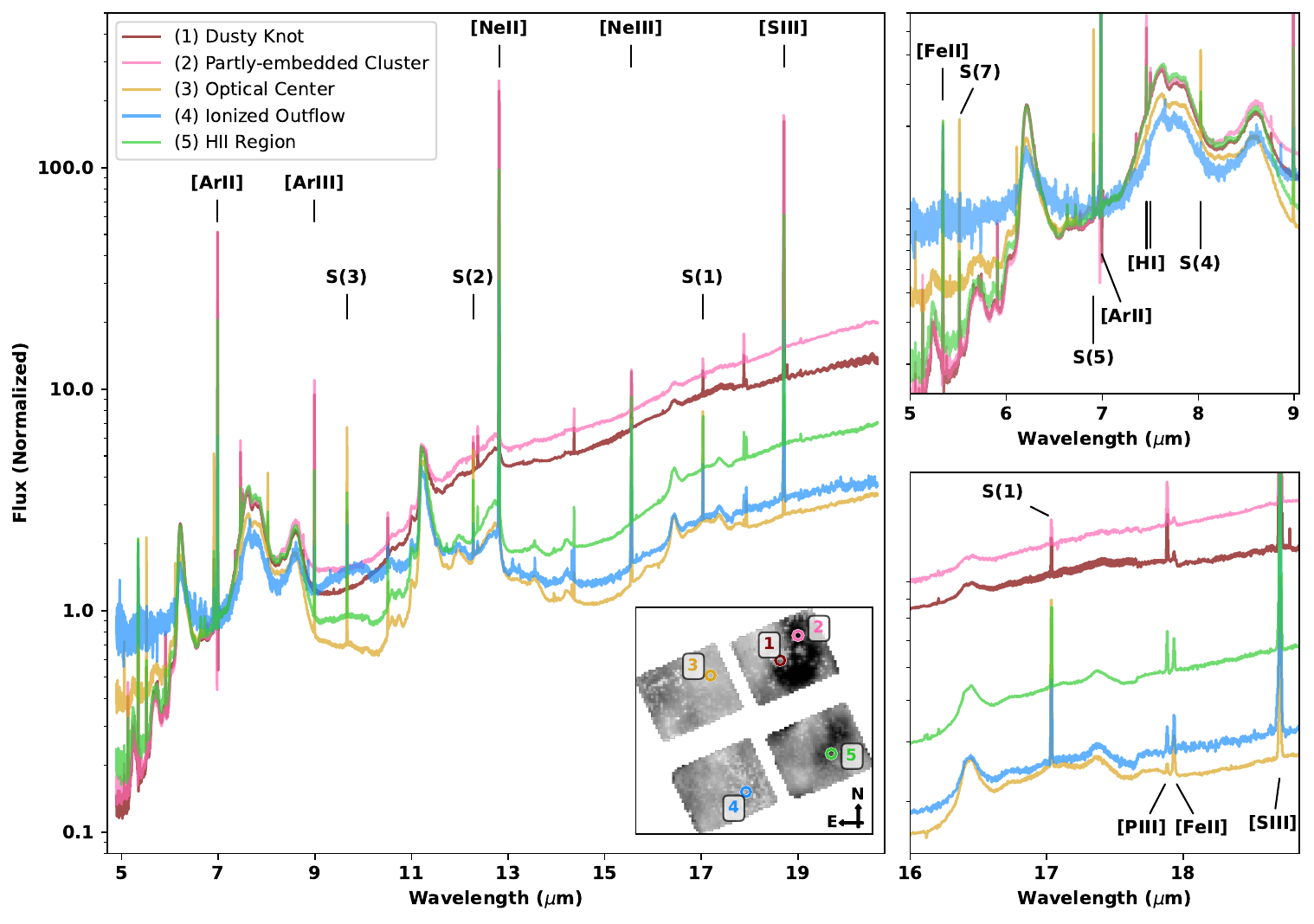} 
\caption{
Aperture-extracted spectra (with no attenuation corrections) of five representative types of environments observed in our MRS observations. In all panels spectra have been normalized to the flux at 7.15 \micron.
{\em Left:} Full spectra (up to 20.7 \micron) for (1) the dusty knot in Region 2 (brown); (2) the partially-embedded cluster in Region 2 (pink); (3) the optical center (yellow); (4) the center of the ionizing outflow in Region 3 (blue); (5) an H{\sc ii} region near the string of YMCs in Region 4 (green). The inset image is the $\tau_{warm}$ map shown in Figure \ref{fig:cafetau} and includes markers showing the location of the apertures ($r$ = 2 pix or 0\farcs26). 
{\em Top Right:} Zoom-in to the 5 -- 9 \micron{} range, which showcases differences in continuum slope at $\sim$7 \micron{} and in relative strength of emission lines and the 6.2 and 7.7 \micron{} PAH complexes.
{\em Bottom Right:} Zoom-in to the 16 -- 19 \micron{} range, which showcases differences in the relative strength of the 17 \micron{} PAH complex and of the \molh{} S(1), [P{\sc iii}] 17.88 \micron, and [Fe{\sc ii}] 17.94 \micron{} lines.
\looseness = -2
}
\label{fig:interestingplaces}
\end{figure*}

To further illustrate our confirmation of the active star-forming ringlet, we extracted spectra within $r = 2$ pixel (0\farcs26) apertures centered on five locations of interest which illustrate the breadth of environments in our FOV. These include (1) a knot in Region 2 with a continuum dominated by cool dust; 
(2) a modestly reddened, likely partially-embedded star cluster in Region 2;
(3) the optical nucleus of M83; (4) the center of the \neii{} outflow in Region 3; and (5) an H{\sc ii} region near the string of YMCs in Region 4. We show spectra for these five locations without any attenuation corrections in Figure \ref{fig:interestingplaces},
with the inset image showing the locations and labels of the extraction apertures overlaid on our map of $\tau_{warm}$. The spectra have been normalized to the flux in a line-free region of the spectrum around $\sim$7.15 \micron{} for easy comparison of the relative continuum shapes and feature strengths. Although we defer a detailed analysis of PAHs and the ionized gas in the core of M83 to future works (Jones et al.\ in prep.), we briefly highlight spectral features which support our interpretation of the dust and molecular gas in these five environments.

The broad dust properties, for example, clearly vary strongly at the scale of $\sim$100 pc, as can be seen by the variety of mid-IR slopes in the left panel of Figure \ref{fig:interestingplaces}. Spectra (1) and (2), which are respectively drawn from a highly- and modestly- attenuated part of our FOV, show deep absorption troughs at 9.7 \micron{} due to silicates, as well as a substantial amount of flux at the red end of our spectra. This suggests the presence of cooler dust and ISM, which is corroborated by the bright CO(2-1) emission (Figure \ref{fig:nucmulti}(d)) and by the low best-fit temperatures inferred by the \ts{} model (bottom panel of Figure \ref{fig:h2slopes}) in the same location. The relative strengths of the neighboring [P{\sc iii}] 17.88 \micron{} (ionizing potential IP = 19.7 eV, tracing moderate ionization zones) and [Fe{\sc ii}] 17.94 \micron{} lines (IP = 7.9 eV, tracing low-ionization zones and/or the shock destruction of iron-rich dust grains; see \citealt{Savage96}) also vary with location, with strong [P{\sc iii}] and weak [Fe{\sc ii}] in environments which are cooler, dustier, and host less-evolved stellar populations (lower-right panel of Figure \ref{fig:interestingplaces}).

In contrast, the [P{\sc iii}]/[Fe{\sc ii}] line ratio approaches unity and then inverts as the eye travels clockwise through our FOV, with comparable line strengths in the more typical H{\sc ii} region and dominant [Fe{\sc ii}] in the high-excitation nucleus and outflow region. Additionally, spectrum (4) is nearly flat across the broad wavelength range covered by our MRS spectroscopy with minimal absorption at 9.7 \micron, suggesting little to no dust in Region 3 \citep{cafe}. Indeed, an examination of archival optical imaging (Figure \ref{fig:nucmulti}(e)) shows a virtually dustless window through the ISM of M83 coincident with the patch of blueshifted and highly-dispersed \neii{} gas we identify in this work. Finally, the optical nucleus and outflow spectra (3 and 4; yellow and blue curves in Figure \ref{fig:interestingplaces}) also show weak but mildly broadened [Ne {\sc v}] emission at 14.3 \micron, which is absent in the other spectra; spectrum 3 additionally shows emission from the very high ionization [Ne {\sc vi}] 7.65 \micron{} line.
We explore the physical drivers of the [Ne {\sc v}] and [Ne {\sc vi}] emission (e.g., fast radiative shocks, X-ray binaries, etc.) in a forthcoming work (Hernandez et al., in prep.).

In H23, we discussed the distribution of iron (as traced by the [Fe{\sc ii}] 25.99 \micron{} line) and its correspondence to soft X-ray emission, which is thought to be driven by both SNRs and strong winds from young, massive stars \citep{Long14}. We found that the [Fe{\sc ii}] emission was generally similar in structure to that of soft X-rays, but had little overall correspondence to the locations of known SNRs. This led us to consider the framework described in the review by \citet{Zhang18}, in which hot starburst-driven galactic winds collide with the cooler, surrounding ISM to induce shocks. These shocks would then produce both soft X-rays and diffuse [Fe{\sc ii}] emission (stemming from the destruction of dust grains that release depleted iron back into the ISM).

As in H23, we emphasize that unambiguous confirmation of this scenario in M83's nuclear region will require self-consistent modeling of the photo- vs.\ shock ionization, which will be presented in an upcoming work (Jones et al.\ in prep.). However, the combination of features discussed above for the high-excitation outflow spectrum -- the absence of $<$5 Myr old YMCs, the absence of dust seen in both optical imaging and our MRS spectroscopy, a low [P{\sc iii}]/[Fe{\sc ii}] line ratio, broadened and blueshifted \neii{} emission, and the presence of [Ne{\sc v}] (IP = 97.12 eV) -- paint a picture in which previous episodes of massive star formation induce shocks and feedback onto the surrounding ISM.

\section{Conclusions}

We used MRS spectroscopy from {\em JWST}/MIRI to examine the warm molecular gas content and star-formation activity in the nuclear region of M83 on the scale of $\sim$3 pc per spaxel. Our resolved analyses illustrates the importance of high-sensitivity spectroscopic observations with {\em JWST} to accurately characterize the multi-phase ISM, the formation of massive star clusters, and the effects of feedback at the smallest scales. We summarize our findings as
follows:

\begin{enumerate}
    \item Our continuum and PAH modeling with {\tt CAFE} shows a remarkable variation in the temperatures and optical depths of dust across our $\sim200\times200$ pc$^2$ field of view. In regions with high dust columns, continuum and line fluxes can be attenuated by non-negligible amounts (up to 30\%) even at the long wavelengths accessible to MRS (Section \ref{subsec:methods:att}).
    
    \item Pure rotational transitions of \molh{} show a range of intensities and excitation conditions within M83's nuclear region, with the most excited emission coincident with the optical nucleus and with groupings of young, massive star clusters. Maps of the \molh{} emission and line widths reveal complex filamentary and clumpy structures (Section \ref{sec:analysis}).

    \item We identify a point-like source in the flux and kinematic maps of the H$_2$ S(1) line, which we propose is a forming or fully-embedded star cluster given its location with respect to the CO (2-1) emission and apparent size ($\lesssim$5 pc) (Section \ref{subsec:disc:baby}).
    
    \item Using the power-law temperature distribution model of \citet{Togi16}, we derive \molh{} masses down to various temperature limits. The warmest \molh{} gas ($>$200 K), totaling 2.26 $\times 10^5 M_{\odot}$, accounts for $\sim$1\% of the total molecular gas mass in our FOV as inferred from CO observations \citep{Hirota18}, while integrating the \ts{} power-law models to temperatures $T_{cal} \sim$57 -- 70 K provide a reasonable match to the CO-based mass estimates. Integrating to still-lower temperatures drastically increases the mass of \molh{} to over 400\% of the CO-based mass estimates, suggesting either the presence of a significant reservoir of gas that is not traced by CO or the need for further constraints on $\alpha_{\mathrm{CO}}$ even in the super-solar metallicity environment of M83's center. Alternative estimates of the molecular gas mass \citep[using, e.g., C{\sc i} far-IR lines][]{Papado22, Ramambason24} will be critical for confirming either scenario (Sections \ref{subsec:analysis:h2mass} and \ref{subsec:disc:codark}).
    
    \item By performing the same analysis on our spaxel-level data and on the integrated-light spectrum of the nuclear region, we find hints that, within the \citet{Togi16} framework, the inferred \molh{} mass may depend weakly on the physical scale probed ($<$5 vs.\ 200 pc) due to blending of high and low-excitation regions (Section \ref{subsec:analysis:h2mass}).
    
    \item The \neii{} 12.81 \micron{} line shows evidence for a secondary, high-FWHM kinematic component that is ubiquitous across our FOV, with a patch of our Region 3 showing evidence for ionized outflows. Converting the \neii{} and \neiii{} luminosities to SFR with diagnostic of \citet{Zhuang19} reveal modest SFR surface densities in Regions 1, 3, and 4, but fairly high ones in Region 2, reaching up to 75 $M_{\odot}$ yr$^{-1}$ kpc$^{-2}$ in $3\times3$ pc$^2$ pixels (Section \ref{subsec:analysis:sfr}).
    
    \item We model the integrated far-UV light from four {\em HST}/COS pointings centered on groupings of YMCs, finding that the youngest and most massive clusters coincide with regions of high SFR and high \molh{} excitation, and are adjacent to extremely dusty, CO-bright clouds. When placed in the context of the high-resolution, multi-wavelength ancillary data available for M83's nuclear region, this supports the notion of a star-forming ringlet with a mild age gradient as first proposed by \citet{Harris01} (Section \ref{subsec:disc:stararc}).
\end{enumerate}

\acknowledgments

{\it Software:} This work makes use of
{\tt CAFE} \citep[Diaz-Santos et al., in prep]{cafe}, {\tt SciPy} \citep{Scipy20}, {\tt numpy} \citep{Numpy11}, and Astropy \citep{Astropy13,Astropy18,Astropy22} in our analyses, as well as {\tt Matplotlib} \citep{Matplotlib07} and {\tt cmasher} \citep{cmasher} for data visualization and {\tt Jupyter} \citep{JupyterNotebook16} for general workflow. 
S.\ H.\ 
acknowledges support from the European Space Agency (ESA). L.\ R.\ gratefully acknowledges funding from the DFG through an Emmy Noether Research Group (grant No. CH2137/1-1).

\bibliography{galstarbib,ismbib,software,m83}

\appendix
\section{Single and double Gaussian fits to \neiii}
\label{sec:appendix}
\restartappendixnumbering

\begin{figure*}[bt]
\centering
\includegraphics[width=0.95\linewidth,trim={0.2cm, 0.1cm, 0.2cm, 0.1cm}, clip]{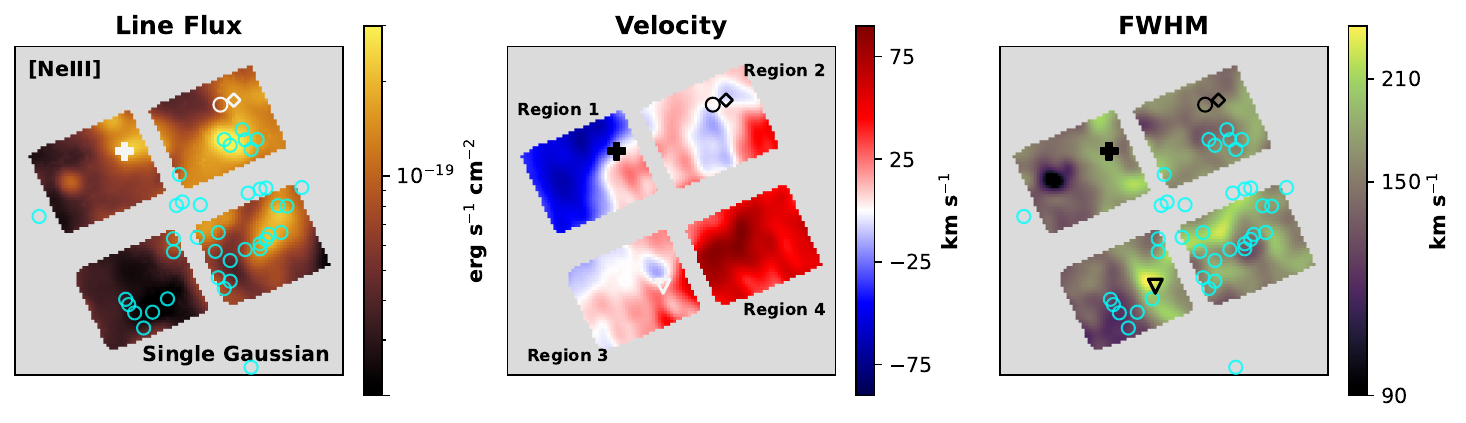}\\
\vspace{20pt}
\includegraphics[width=0.95\linewidth,trim={0.2cm, 0.1cm, 0.2cm, 0.1cm}, clip]{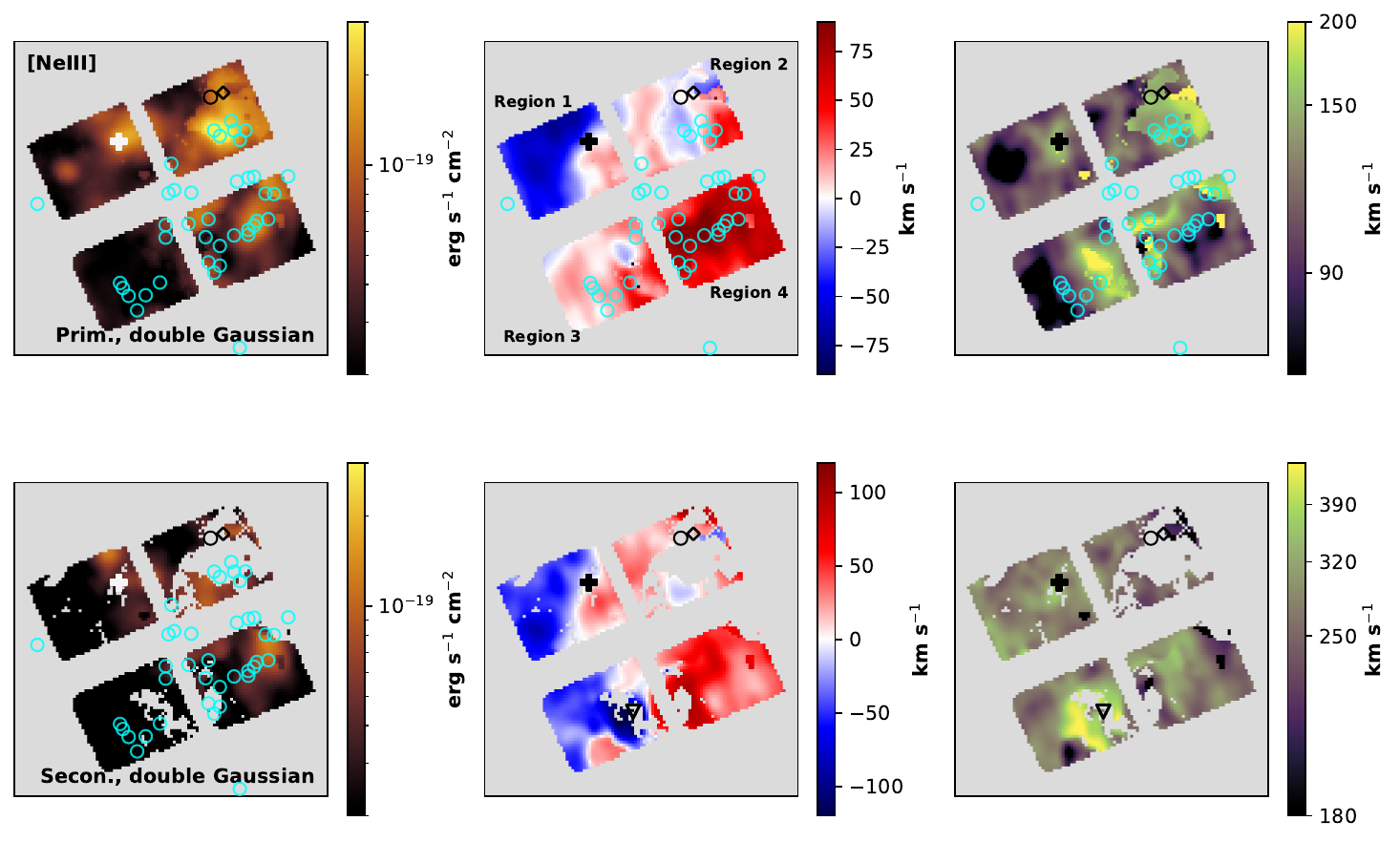}
\caption{
Flux, velocity shift, and FWHM of the \neiii~15.5\micron{} line using either a single Gaussian fit ({\em top row}) or a two-component Gaussian fit ({\em middle and bottom}) after application of the statistical $F$-test and our S/N cut. Cyan circles mark the position of UV-bright star clusters, and other symbols are as defined in Figure \ref{fig:h2slopes}. 
\looseness = -2
}
\label{fig:appendix:neon3}
\end{figure*}

In Figure \ref{fig:appendix:neon3}, we show the flux, velocity shift, and FWHM of the \neiii{} 15.5 \micron{} line using both a single and a double Gaussian component. Details on the line fitting and on the application of the statistical $F$-test for the two-component fits are given in Sections \ref{subsec:methods:emlines} and \ref{subsec:analysis:sfr}, respectively.

\end{document}